\documentclass{aastex63}
\usepackage{graphicx}

\usepackage{natbib}
\usepackage{mathrsfs}
\usepackage{amsmath}
\usepackage{array}
\usepackage{multirow}
\newcolumntype{P}[1]{>{\centering\arraybackslash}p{#1}}

\shorttitle{Failed solar eruption of torus unstable flux rope}
\shortauthors{Mitra et al.}

\begin{document}

\title{Multiwavelength signatures of episodic null point reconnection in a quadrupolar magnetic configuration and the cause of failed flux rope eruption}

\correspondingauthor{Prabir K. Mitra}
\email{prabir@prl.res.in}

\author[0000-0002-0341-7886]{Prabir K. Mitra}
\affiliation{Udaipur Solar Observatory, Physical Research Laboratory, Udaipur 313 001, India}
\affiliation{Department of Physics, Gujarat University, Ahmedabad 380 009, India}
\affiliation{IGAM/Institute of Physics, University of Graz, Universit\"{a}tsplatz 5, A-8010 Graz, Austria}

\author[0000-0001-5042-2170]{Bhuwan Joshi}
\affiliation{Udaipur Solar Observatory, Physical Research Laboratory, Udaipur 313 001, India}

\author[0000-0003-2073-002X]{Astrid M. Veronig}
\affiliation{Institute of Physics \& Kanzelh\"{o}he Observatory, University of Graz, Universit\"{a}tsplatz 5, A-8010 Graz, Austria}

\author[0000-0001-6238-0721]{Thomas Wiegelmann}
\affiliation{Max-Planck-Institut f\"{u}r Sonnensystemforschung, Justus-von-Liebig-Weg 3, D-37077 G\"{o}ttingen, Germany}

\begin{abstract}
In this paper, we present multiwavelength observations of the triggering of a failed-eruptive M-class flare from the active region NOAA 11302, and investigate the possible reasons for the associated failed eruption. Photospheric observations and Non-Linear Force Free Field extrapolated coronal magnetic field revealed that the flaring region had a complex quadrupolar configuration with a pre-existing coronal null point situated above the core field. Prior to the onset of the M-class flare, we observed multiple periods of small-scale flux enhancements in GOES and RHESSI soft X-ray observations, from the location of the null point. The pre-flare configuration and evolution reported here are similar to the ones presented in the breakout model but at much lower coronal heights. The core of the flaring region was characterized by the presence of two flux ropes in a double-decker configuration. During the impulsive phase of the flare, one of the two flux ropes initially started erupting but resulted in a failed eruption. Calculation of the magnetic decay index revealed a saddle-like profile where decay index initially increased to the torus unstable limits within the heights of the flux ropes but then decreased rapidly reaching to negative values, which was most likely responsible for the failed eruption of the initially torus unstable flux rope. 
\end{abstract}



\section{Introduction} \label{Sec_Introduction} 
Solar eruptive phenomena are violent activities occurring in the solar atmosphere which include catastrophic energy release within a short time in a localized region i.e., flares \citep{Priest2002, Fletcher2011, Benz2017} along with the expulsion of plasma and magnetic field into the interplanetary space i.e., coronal mass ejections \citep[CMEs;][]{Chen2011, Green2018}. Earth-directed CMEs inflict hazardous effects in the near-Earth environment that include damage in the satellites, disruption of the telecommunication system, damage in the electrical power grids on Earth etc. \citep[space weather; see,][]{Moldwin2008, Koskinen2017, Lanzerotti2017}. While most of the major flares are associated with CMEs (i.e., eruptive flares), a significant number of flares do not lead to CMEs \citep[i.e., confined flares; see e.g.,][]{Yashiro2005, Baumgartner2018, Li2020}. Observationally, a special variant of confined flares also involve `failed flux rope eruption' where a flux rope initially gets activated from the source region but subsequently fails to escape from the overlying layers of solar corona and eventually the material falls back \citep[][see also \citealp{Ji2003, Alexander2006, Liu2009, Kushwaha2015}]{Gilbert2007}. With an ever-increasing urge to understand the factors leading to CME eruptions and develop methods to predict space-weather, the observational and theoretical studies on failed eruptions have recently gained much attention and have become an important research topic in contemporary solar physics \citep[see e.g.,][]{Cheng2011, Sarkar2018, Amari2018}.

Eruptive flares are usually characterized by the formation of two parallel, ribbon-like brightenings followed by the development of a coronal flare-arcade connecting the two ribbons \citep{Svestka1992}. To explain such two-ribbon flares, the `standard flare model', also known as the `CSHKP' model was developed that includes the pioneering works of \citet{Carmichael1964, Sturrock1966, Hirayama1974, Kopp1976}. In recent years, this model has been extended to three-dimension \citep{Aulanier2012, Aulanier2013, Janvier2013}. According to this model, when a pre-existing magnetic flux rope (MFR) is triggered for upward eruption, inflow of magnetic field takes place beneath the MFR where eventually a current sheet is generated. Magnetic reconnection at such current sheets rapidly convert previously stored magnetic energy into plasma heating and kinetic energy of accelerated particles, and adds also magnetic flux to the erupting flux rope prolonging the driving Lorentz force \citep{Vrsnak2016, Veronig2018}. The CSHKP model is quite successful in explaining commonly observed large-scale features of eruptive flares e.g., footpoint and looptop hard X-ray (HXR) sources; the expansion of flare ribbons in opposite polarities observed in optical, EUV and SXR wavelengths; formation of highly structured loop arcades observed in optical, EUV and SXR wavelengths, that connect the flare ribbons; formation of hot cusp like structures; etc \citep[see e.g.,][]{Tsuneta1992, Masuda1994, Veronig2006, Sui2006, Miklenic2007, Joshi2017, Guo2019, Mitra2019}. However, some important aspects related to solar eruptions remain out of scope the standard flare model, e.g., the processes of flux rope formation; onset/triggering of eruptions; the initial dynamical evolution of CME/flux rope; etc \citep[][see also \citealp{Joshi2019}]{Green2018}.

Theoretically an MFR is defined as a set of magnetic field lines which are wrapped around a central axis \citep{Gibson2006}. Observationally, MFRs can be identified in the form of filaments/prominences, coronal cavities, coronal sigmoids, hot coronal channels etc \citep[see review by,][]{Patsourakos2020}. Filaments are dark, thread-like structures observed in the chromospheric images of the Sun \citep{Martin1998}. When filaments are observed above the limb of the Sun, they appear as bright structures and therefore, they are called prominences \citep[see review by,][]{Gibson2018}. Although, the exact structure of filaments are still debatable \citep[see e.g.,][]{Antiochos1994}, it is now believed that MFRs form the basic structures of the active region filaments. Coronal cavities represent the transverse cross-sectional view of MFRs, where the accumulation of plasma can be observed at the bottom of the dark cavity providing important insights on the relation between the MFRs and filaments \citep{Gibson2015}. Coronal sigmoids are `S' (or reverse `S') shaped structures observed in SXR \citep{Rust1996, Manoharan1996} and EUV wavelengths \citep{Joshi2017, Mitra2018}, and are interpreted as manifestations of highly twisted MFRs \citep{Green2018}. Hot channels are coherent structures observed in the high temperature pass-band EUV images of the solar corona, which indicate activated, quasi-stable MFRs \citep{Zhang2012, Cheng2013, Nindos2015, Joshi2018, Hernandez2019, Mitra2019, Sahu2020, Kharayat2021}. Two possible scenario of flux rope formation have been proposed: emergence of MFRs from the convection zone of the Sun to the solar atmosphere by magnetic buoyancy \citep{Archontis2008, Chatterjee2013} and formation of MFRs from sheared arcades, in response to small-scale magnetic reconnection in the corona \citep{Aulanier2010, Inoue2018, Mitra2020}.

Successful eruption of MFRs are essential for the generation of CMEs. In order to explain the triggering of MFRs toward eruption, different models have been proposed which can be largely classified into two groups: models relying on ideal MHD instability (kink and torus instability) and models relying on magnetic reconnection (tether-cutting model, breakout model etc.). According to the torus instability \citep{Torok2006}, a toroidal current ring may be triggered to erupt if the ambient poloidal magnetic field decreases with height faster than a critical rate. The magnetic decay index ($n$), defined as $n=-\frac{d \log(B_{p})}{d \log(z)}$; where $B_{p}$ and $z$ are the external poloidal magnetic field and height, respectively, is used as the quantification of the decay of magnetic field with height above the relevant polarity inversion line. Theoretically it was found that an MFR is subject to torus instability when it reaches a region characterized by $n>1.5$ \citep{Bateman1978}. However, a number of observational studies revealed that the threshold value of $n$ for torus instability, lies within the range [1.1--1.75] \citep[e.g.,][]{Liu2008, Kliem2013, Zuccarello2015}. Kink instability suggests that an MFR may be destabilized if its twist increases beyond $\approx$3.5$\pi$ \citep{Torok2004}. The tether-cutting model explains triggering of solar eruptions from a highly sheared bipolar magnetic configuration where the initial magnetic reconnection takes place deep within the sheared core field \citep{Moore1992}. Here, the flux rope is developed from the sheared arcades in response to these pre-flare magnetic reconnections and the successful eruption of the flux rope depends on the flux content of the sheared core field relative to the overlying envelope field \citep{Moore2001}. Contrary to the tether-cutting model, the breakout model involves complex multipolar topology with one or more null points situated well above the core field \citep{Antiochos1999}. As the photospheric magnetic field evolves (in the form of flux emergence, shearing motion, etc.), the core field extends outward stretching the null point leading to the formation of breakout current sheets \citep{Karpen2012}. The initial reconnection (breakout reconnection) takes place on these breakout current sheets, which results in reducing the downward strapping force of the enveloping field and allowing the core field to erupt successfully. Notably, once the triggered flux rope attains upward eruptive motion, standard flare reconnection (as explained in the CSHKP model) sets in beneath the flux rope, a process which is common for all the triggering mechanisms.

The presence of the coronal null point in the pre-flare configuration is an essential requirement in the breakout model as the breakout current sheet is formed by stretching the null point. Theoretically, magnetic null points are defined as locations where all the three components of magnetic field become zero and increase linearly with distance from it \citep[Chapter 6 in][]{Priest2014}. Thus, null points create discontinuities in the coronal magnetic field separating different domains of flux regions. In general, regions of strong gradients in continuous magnetic fields are identified as quasi-separatrix layers \citep[QSLs;][]{Priest1995}. The gradient of magnetic connectivity can be quantified by computing the degree of `squashing factor' ($Q$) by calculating the norm ($N$) of the field line mapping of magnetic domains \citep{Demoulin1996, Pariat2012}. Theoretically, \citet{Titov2002} showed that in all physical scenarios, $Q=2$ is the lowest possible value of $Q$ and QSLs are, therefore, characterized by $Q\gg2$ \citep{Aulanier2005} while $Q\rightarrow\infty$ is representative of null points. Due to local diffusion, magnetic fields in QSLs can constantly change their connectivities \citep{Aulanier2006} which observationally can be identified as apparent slipping motion of flare kernels in imagery of the lower solar atmosphere, termed as `slipping' or `slip-running' reconnection \citep[see e.g.,][]{Janvier2013}.

In this paper, we report on an M4.0 flare from active region NOAA 11302 on 2011 September 26, which was associated with the failed eruption of a filament. Different multi-wavelength aspects of this event with the main focus on the non-thermal energy evolution, was studied by \citet{Kushwaha2014}. Their observations suggest that the impulsive phase of the flare was characterized by two short-lived microwave (MW) peaks in 17 and 34 GHz observed by the Nobeyama Radioheliograph \citep[NoRH;][]{Nakajima1994, Takano1997}. Interestingly, while the first MW burst occurred co-temporal with a sudden peak in hard X-ray (HXR) wavelengths observed by the \textit{Reuven Ramaty High Energy Solar Spectroscopic Imager} \citep[\textit{RHESSI};][]{Lin2002} of energies up to $\approx$200 keV following a power law with a hard photon spectral index ($\gamma$) $\sim$3, the second non-thermal burst observed in MW was much less pronounced in HXRs. Importantly, the onset of the X-ray emission during the flare occurred immediately after the emergence of a pair of small-scale magnetic transients of opposite polarities at the inner core region which led them to conclude that small-scale changes in the magnetic structure may play a crucial role in triggering the flare process by disturbing the pre-flare magnetic configurations. In view of the rapid temporal evolution of the HXR and MW flux during the early impulsive phase of the flare, they concluded that an abrupt energy release via spontaneous magnetic reconnection was responsible for the occurrence of the flare. In order to investigate whether this apparent spontaneous magnetic reconnection was influenced by topological features e.g., flux ropes, null points, etc., we revisited the event considering a longer period of observation and analysis. Our study readily revealed that the flare was preceded by a number of subtle SXR flux enhancements suggesting a possible influence of external factors to be responsible for the triggering of the flare. By employing a Non-Linear Force Free Field (NLFFF) extrapolation model, complemented by EUV observations of the \textit{Atmospheric Imaging Assembly} \citep[AIA;][]{Lemen2012} on board the \textit{Solar Dynamics Observatory} \citep[SDO;][]{Pesnell2012}, we investigate the causal connection between the activities during the extended preflare phase and the main flare, and investigate the factors responsible for the failed filament eruption. The structure of the paper is as follows: Section \ref{Sec_Data} provides a brief description of the observational data sets and different analysis techniques used in this work. We discuss the temporal evolution of X-ray and EUV emission from the flaring active region in Section \ref{Sec_Lightcurve}. Section \ref{Sec_Preflare_conf} gives a detailed overview of the photospheric magnetic configuration of the active region as well as discusses the coronal magnetic configuration of the flaring region. Section \ref{Sec_Results} presents the observational results on the basis of coronal and chromospheric imaging analysis. The results obtained by modeling of coronal magnetic field are explained in Section \ref{Sec_Extrapolation}. We discuss and interpret the results in Section \ref{Sec_Discussion}.

\section{Observational Data and Analysis Techniques} \label{Sec_Data}
\subsection{Data} \label{Sec_source}
In this work, we have combined multi-wavelength observational data from different sources. For EUV imaging of the Sun, we utilized observations provided by the AIA on board SDO. In particular, we extensively used the full disk 4096$\times$4096 pixel solar images in 304 \AA\ ($\log(T)=4.7$), 171 \AA\ ($\log(T)=5.8$), 94 \AA\ ($\log(T)=6.8$) and 131 \AA\ ($\log(T)=5.6, 7.0$) channels at a pixel scale of 0$\farcs$6 and a cadence of 12 s. In order to enhance the fine structures, all the AIA images have been filtered with the `unsharp\_mask' image processing algorithm. Chromospheric observations of the Sun in the H$\alpha$ passband were obtained from the Global Oscillation Network Group \citep[GONG;][]{Harvey1996}. GONG provides 2096$\times$2096 pixel full disk H$\alpha$ images\footnote{See \url{http://halpha.nso.edu}.} of the Sun with a pixel scale of $\approx$1$\farcs$0 \citep{harvey2011}. To study the evolution of photospheric magnetic structures, we used data from the \textit{Helioseismic and Magnetic Imager} \citep[HMI;][]{Schou2012} on board SDO. Among the different products of HMI, we have utilized the 4096$\times$4096 pixel full disk intensity images and line-of-sight (LOS) magnetograms at a pixel scale of 0$\farcs$5 and cadence of 45 s. We used X-ray observations of the Reuven Ramaty High Energy Solar Spectroscopic Imager \citep[RHESSI;][]{Lin2002} which has a spatial resolution (as fine as $\sim$2$\farcs$3) and energy resolution (1--5 keV) over the energy range 3 keV--17 MeV. For constructing RHESSI X-ray images, we used the PIXON algorithm \citep{Metcalf1996} with the natural weighting scheme for front detector segments 2--9 (excluding 7).

\subsection{Numerical Analysis Techniques} \label{Sec_tech}
In order to investigate the coronal magnetic configuration of the active region with the aim to understand the cause of the failed eruption associated with the event, we carried out coronal magnetic field modeling using the optimization based non-linear force free field (NLFFF) extrapolation technique developed by \citet{Wiegelmann2010, Wiegelmann2012}. As the boundary condition for the extrapolation, we used the vector magnetogram at 2011 September 26 04:58 UT from the ``hmi\_sharp\_cea\_720s" series of HMI/SDO with a reduced resolution of 1$\farcs$0 pixel$^{-1}$. The extrapolation was done in a Cartesian volume of 616$\times$288$\times$256 pixels which corresponds to physical dimensions of $\approx$447$\times$209$\times$186 Mm$^3$. According to the theory of the NLFFF model, the angle between current ($\vec{J}$) and magnetic field ($\vec{B}$) should be 0 as the total value of the Lorentz force i.e., $|\vec{J}\times\vec{B}|$ is 0. However, since the approach of extrapolation is based on numerical techniques, the NLFFF-reconstructed magnetic field is expected to return non-zero values of $|\vec{J}\times\vec{B}|$. Therefore, to assess the quality of the coronal magnetic field reconstruction, the average value of the fractional flux ratio ($|f_i|=|(\vec{\nabla}\cdot\vec{B})_i|/(6|\vec{B}|_i/{\bigtriangleup}x))$, weighted angle ($\theta_J$) between $\vec{J}$ and $\vec{B}$ can be considered \citep[see,][]{DeRosa2015}. In this study, we obtained the values of the residual errors by averaging these parameters over the entire computation domain i.e., 616$\times$288$\times$256 pixels with pixel dimensions physically translating to $\approx$0.725 Mm. The residual errors were found to be:

\begin{equation} \label{Eq_assess_NLFFF}
<|f_i|>\approx4.06\times10^{-4}; ~~~~~ \frac{|\vec{J}\times\vec{B}|}{|\vec{J}\cdot\vec{B}|}\approx0.14; ~~~~~ \theta_J\approx6.^\circ76.
\end{equation}

\noindent {In general, NLFFF solutions are considered as good solutions if they return the values: $<|f_i|>\lesssim2\times10^{-3}$ and $\theta_J\lesssim10^\circ$ \citep[see, e.g.,][]{DeRosa2015}.

Using the NLFFF extrapolation results, we calculated the degree of squashing factor ($Q$) and the twist number ($T_w$) within the extrapolation volume by employing the code developed by \citet{Liu2016}. In order to locate 3D null-points within the extrapolation volume, we used the trilinear method \citep{Haynes2007}, by dividing the whole active region volume into grid cells of dimension 2$\times$2$\times$2 pixels. If any of the three components of the magnetic field vector have same sign at all the eight corners of the grids, it is considered that the corresponding grid cell cannot contain any null point within it and therefore, the corresponding cell is excluded from further analysis. Each of the remaining other cells is then further divided into 100$\times$100$\times$100 subgrid cells and the null point is located by using the Newton-Raphson method\footnote{\url{http://fourier.eng.hmc.edu/e176/lectures/NM/node21.html}} for finding roots of equations. This iterative method was continued until the uncertainty in the solution reached to $\lesssim$2 subgrid cell width. For visualizing the results obtained from the NLFFF extrapolation, we used the Visualization and Analysis Platform for Ocean, Atmosphere, and Solar Researchers \citep[VAPOR;][]{Clyne2007} software.
 
\section{Temporal Evolution of X-ray and EUV Emission} \label{Sec_Lightcurve}

\begin{figure}
\epsscale{1.2}
\plotone{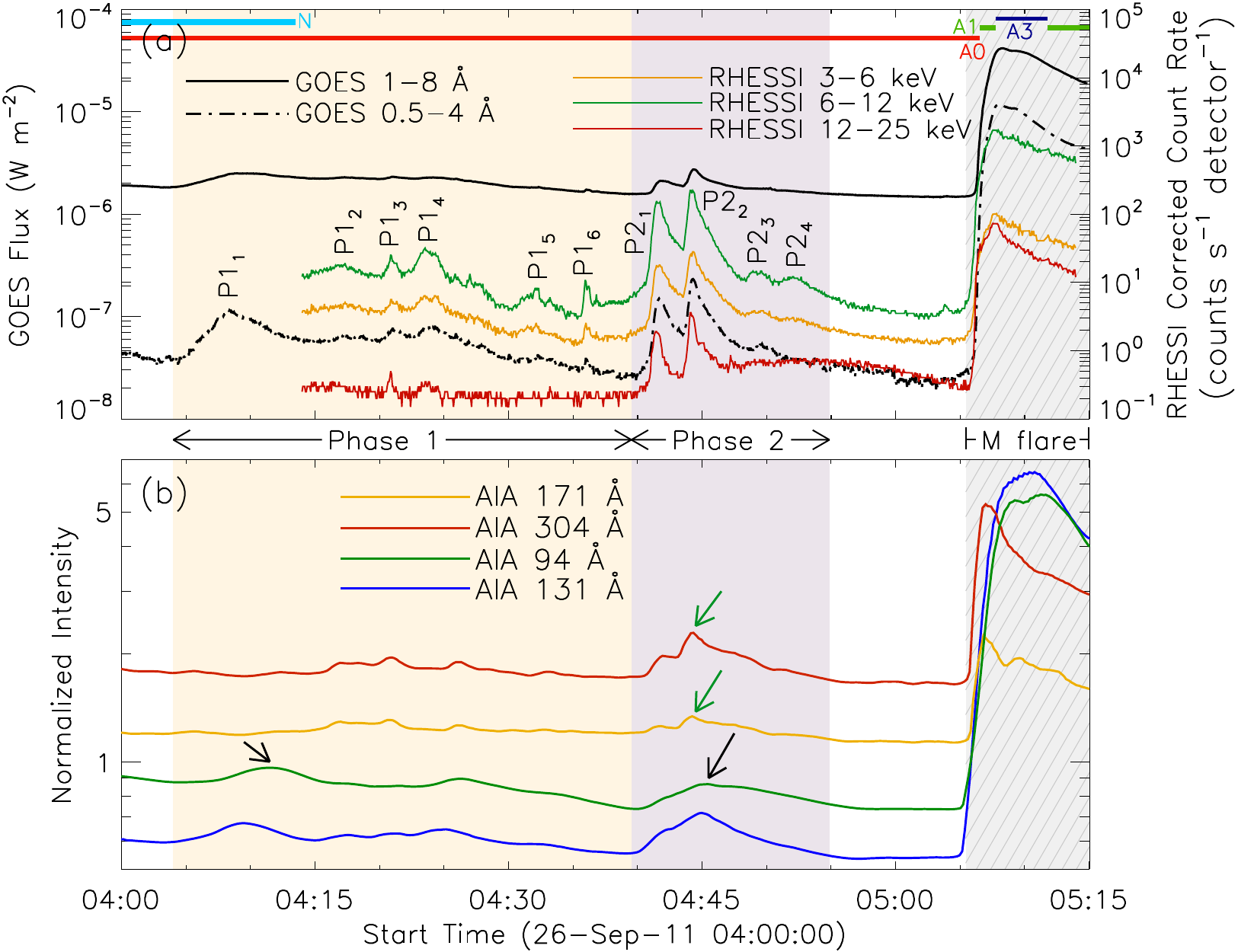}
\caption{Panel (a): Evolution of X-ray flux in different GOES and RHESSI channels showing the extended pre-flare phase along with the evolution of the M4.0 flare. For better visualization, RHESSI fluxes are normalized by $\frac{1}{2}$ and $\frac{1}{20}$ for 3--6 keV and 12--25 keV channels. The horizontal bars at the top represent the status of RHESSI observation and attenuator states. Panel (b): Normalized AIA EUV intensity variations in the same duration as in panel (a). For better visualization, AIA 171, 304, 94 and 131 \AA\ channels are scaled by factors of 1.2, 1.8, 0.9, and 0.6, respectively. The light yellow and light purple shaded intervals represent the durations of the two pre-flare periods of SXR flux enhancements which are further identified as `Phase 1' and `Phase 2', respectively. Individual SXR peaks during Phase 1 and Phase 2 are identified as P1$_1$--P1$_6$ and P2$_1$--P2$_4$, respectively. The hatched light grey-shaded interval represents the impulsive and early gradual phase of the M2.4 flare. The two green arrows in panel (b) indicate the intensity enhancement in AIA 171 and 304 \AA\ channels during `P2$_2$' while the two black arrows indicate mild enhancement in the intensities in AIA 94 and 131 \AA\ channels during P1$_1$ and P2$_2$.}
\label{Fig_Lightcurve}
\end{figure}

In Figure \ref{Fig_Lightcurve}, we compare the temporal evolution of the X-ray fluxes in the 1--8 \AA\ and 0.5--4 \AA\ channels of GOES and multiple energy channels of RHESSI covering the energy range 3--25 keV (Figure \ref{Fig_Lightcurve}(a)) and EUV fluxes derived from different AIA channels (Figure \ref{Fig_Lightcurve}(b)) during 2011 September 26 04:00 UT -- 05:15 UT which included an extended phase prior to the onset of the M4.0 flare as well as the impulsive and a part of the gradual phase of it. GOES did not observe the Sun during 05:15 UT -- 06:25 UT which covered most of the declining phase of the flare\footnote{See \url{https://www.swpc.noaa.gov/products/goes-x-ray-flux}.}. RHESSI missed the initial phase of our observing period (till $\sim$04:13 UT) due to `RHESSI night'. From Figure \ref{Fig_Lightcurve}(a), we find a number of episodes of small emission enhancements prior to the M-class flare during $\approx$04:04 UT -- 04:53 UT, which were most prominent in GOES 0.5--4 \AA\ and RHESSI channels up to 12 keV. Multiwavelength EUV and X-ray imaging (discussed in detail in Sections \ref{Sec_Preflare}) suggested a localized region to undergo compact brightening during these flux enhancing periods. Based on the intensity and compactness of the EUV and X-ray sources, we divided the whole period into two phases: `Phase 1': 04:04 UT -- 04:39 UT (highlighted by light yellow background in Figure \ref{Fig_Lightcurve}) and `Phase 2': 04:39 UT -- 04:53 UT (highlighted by light purple background in Figure \ref{Fig_Lightcurve}) and identify the flux peaks by P1$_1$--P1$_6$ and P2$_1$--P2$_4$, respectively. The onset of the flare occurred at $\approx$05:06 UT which was followed by an impulsive rise of flare emission. The flare reached its peak at $\approx$05:08 UT after undergoing a brief impulsive phase lasting for only $\approx$2 min. AIA EUV intensities displayed a general agreement with the X-ray flux variation although the small-scale enhancements observed in the GOES 0.5--4 \AA\ and RHESSI lightcurves up to energies $\approx$12 keV during Phase 1 was much less pronounced in EUV lightcurves. Notably, the AIA 171 and 304 \AA\ intensities showed a significant enhancement during the P2$_2$ peak (indicated by the green arrows in Figure \ref{Fig_Lightcurve}(b)) while they remained completely unchanged during P1$_1$. Intensities in the AIA 94 and 131 \AA\ channels showed mild enhancements during both the aforementioned X-ray peaks (indicated by the black arrows in Figure \ref{Fig_Lightcurve}(b)).

It is worth mentioning that the soft and hard X-ray lightcurves obtained from GOES and RHESSI (Figure \ref{Fig_Lightcurve}(a)) are disk-integrated, i.e., includes emission originating from the full solar disk. On the other hand, the EUV light curves from SDO/AIA displayed in Figure \ref{Fig_Lightcurve}(b) were derived by integrating the counts within the flaring region of FOV=[(-405$\farcs$:-645$\farcs$),(30$\farcs$:230$\farcs$)], shown in Figure \ref{Fig_Intro}(d). Despite the difference in the regions used to construct the lightcurves in Figures \ref{Fig_Lightcurve}(a) and (b), their general agreement suggests that the overall activity of the Sun was primarily dominated by the flaring activities occurring within the active region NOAA 11302. This is expected since the intensities in the EUV, SXR and HXR domains from flaring active regions increase by orders of magnitude during flares, thus, becoming the dominant contribution to the changes of the full-Sun light curves.

\begin{figure}
\epsscale{0.9}
\plotone{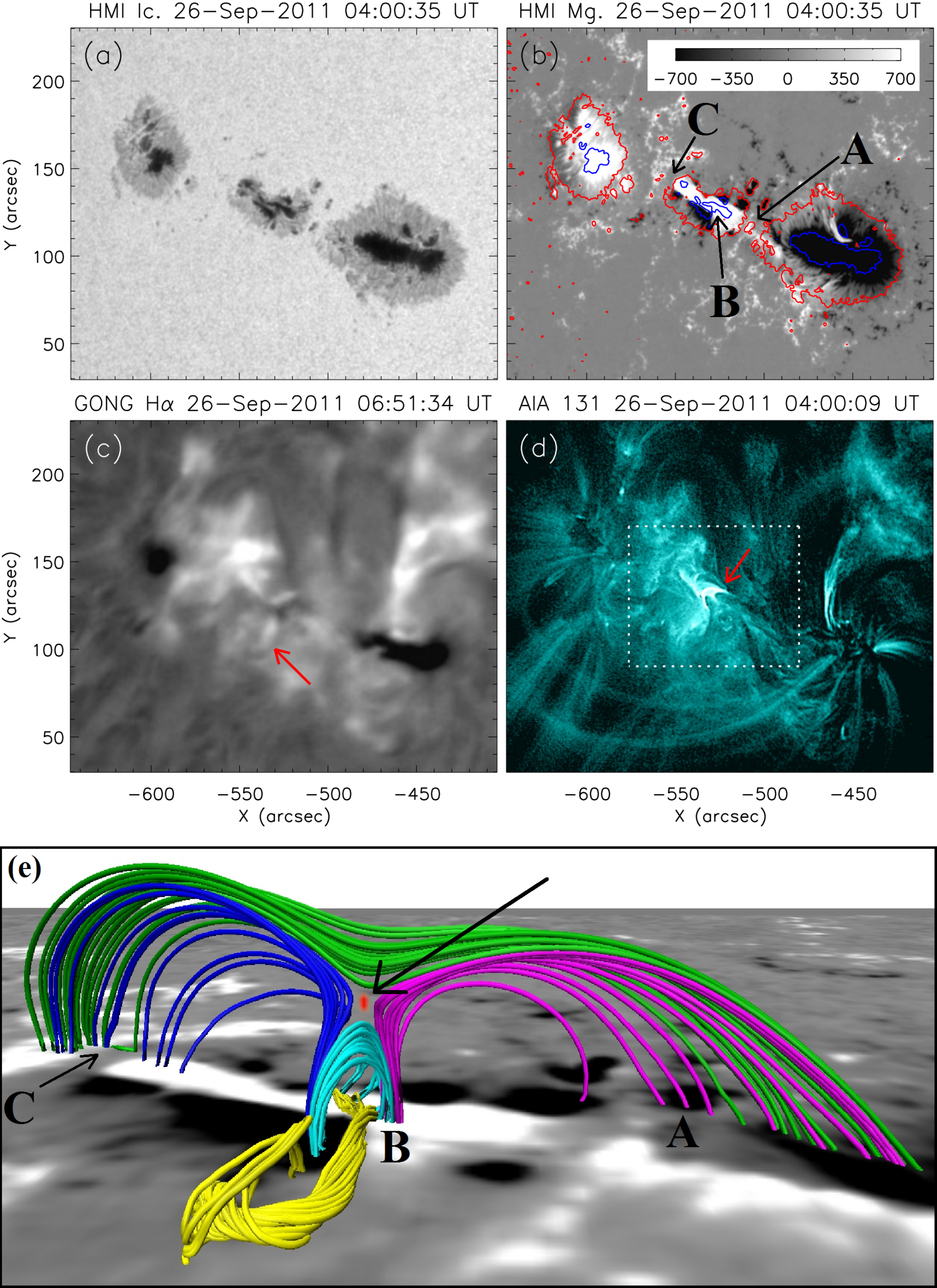}
\caption{Panel (a): HMI white-light image of AR 11302 on 2011 September 26 04:00 UT. Panel (b): Cotemporal HMI line-of-sight (LOS) magnetogram. Magnetic flux density in $Gauss$ is color-coded as the colorbar shown in the inset. The blue and red contours in panel (b) indicates the spatial extent of the umbrae and penumbrae of the sunspot groups as observed in white light. The contour levels are 35\% and 75\% of the maximum white light emission, respectively. Panel (c): GONG Learmonth H$\alpha$ image of the active region showing the presence of multiple filaments. The red arrow indicates a faint filament which was involved in the M-class flare under study. Panel (d): AIA 131 \AA\ image of the active region showing a set of significantly bright, small coronal loops during the pre-flare phase (indicated by the red arrow). The white dotted boxes in panel (d) indicate the field of view of Figures \ref{Fig_GOES_AIA}(b)--(m). Panel (e): NLFFF extrapolation results of the flaring region showing a flux rope (displayed by the yellow lines) and four sets of coronal loops connecting different photospheric flux regions associated with the quadrupolar configuration of the central sunspot group. The red patch (indicated by the black arrow) represent a coronal null point obtained by the trilinear method, which was situated in the quadrupolar configuration. The red patch is further characterized by the $Q$ value of $\log(Q)=8.5$. The stamps A, B, and C in panels (b) and (e) are for comparison between the field of views (FOVs) of the corresponding panels.}
\label{Fig_Intro}
\end{figure}
 
\section{NOAA 11302: $\delta$-Sunspot Region and Quadrupolar Configuration}
\label{Sec_Preflare_conf}

\begin{figure}
\epsscale{1.2}
\plotone{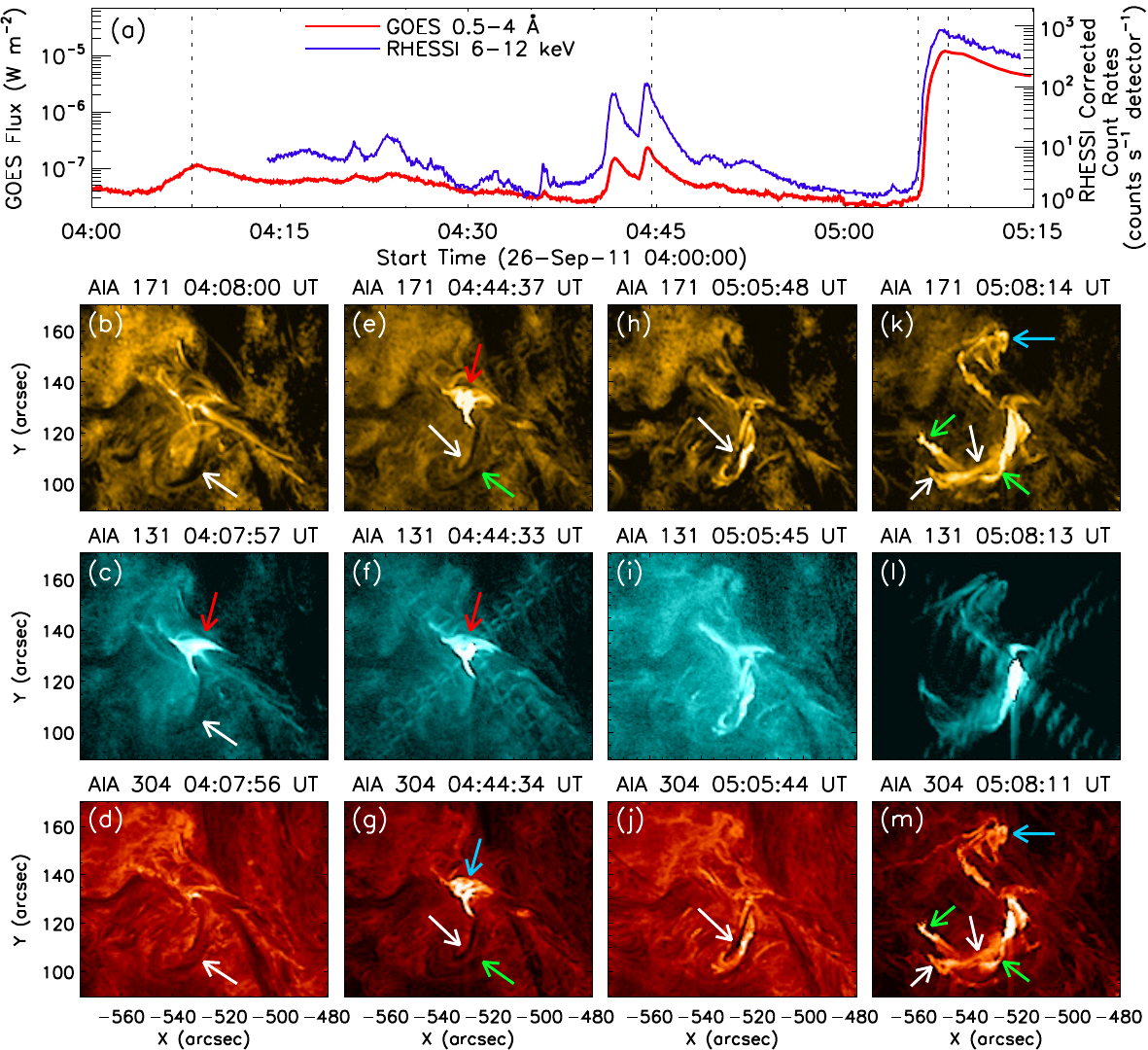}
\caption{Panel (a): SXR flux in the GOES 0.5--4 \AA\ (red curve) and RHESSI 6--12 keV (blue curve) channels. Panels (b)--(m): Multi-wavelength evolution of the core active region in AIA 171 \AA\ (panels (b), (e), (h), (k)), 131 \AA\ (panels (c), (f), (i), (l)) and 304 \AA\ (panels (d), (g), (j), (m)). For comparison, four dashed lines are drawn in panel (a) corresponding to the time-informations of the images of the four columns. The white and green arrows in different panels indicate filaments during different phases within the studied interval. The red arrows in different panels and the sky-colored arrow in panel (g) indicate the location of pre-flare brightening. The sky-colored arrows in panels (k) and (m) indicate a remote brightening observed during the peak phases of the flare. The field-of-view of these panels are shown by the dotted boxes in Figures \ref{Fig_Intro}(d). An animation of the figure is available. The animation has a duration of 23 s and displays the temporal evolution of the active region in AIA 171, 131, and 304 \AA\ wavelengths with 12 s cadence, between 04:00 UT and 05:15 UT on 2011 September 26.}
\label{Fig_GOES_AIA}
\end{figure}

During the M-class flare under study, active region (AR) NOAA 11302 was centered at the heliographic location $\approx$N15E35. The white light image of the AR (Figure \ref{Fig_Intro}(a)) shows that the AR was comprised of three separate sunspot groups which were distributed along the northeast--southwest (NE--SW) direction. Apart from the three prominent sunspots, the AR also contained few small pores. Comparison of a co-temporal magnetogram of the AR (Figure \ref{Fig_Intro}(b)) with the white light image, suggests that the easternmost sunspot group was of predominantly positive polarity; while, the westernmost sunspot group was mostly composed of negative flux although few dispersed positive flux regions were distributed around it. The middle sunspot group is particularly interesting as it had a bipolar magnetic structures. From the intensity contours over the magnetogram in Figure \ref{Fig_Intro}(b), it becomes evident that this sunspot group had fragmented umbrae of opposite polarities within a single penumbra; suggesting the AR to be a $\delta$-type AR. Notably, the M-class flare and associated failed eruption reported in this paper, originated from this sunspot group. The chromospheric H$\alpha$ image shows a very small, faint filament with one leg attached to the middle sunspot group (indicated by the red arrow in Figure \ref{Fig_Intro}(c)). Coronal images of the active region, particularly in the hot AIA EUV channels e.g., 94 \AA\ and 131 \AA , revealed an interesting structure at the northern end of the filament (indicated by the red arrow in Figure \ref{Fig_Intro}(d)), which was composed of multiple small coronal loops connecting different polarity regions of the middle sunspot group.

NLFFF extrapolation results readily validated the observed coronal structures, both at large and small scales. In Figure \ref{Fig_Intro}(e), we show the modeled coronal configuration associated with the central sunspot group. Notably, the coronal loops shown in blue, green, pink and sky-colors, constituted a well-defined quadrupolar coronal configuration. Importantly, our analysis further confirms the presence of a coronal null point in the quadrupolar configuration above the central sunspot group (shown by the red patch in Figure \ref{Fig_Intro}(e); indicated by the black arrow). NLFFF extrapolation also demonstrated the presence of a flux rope associated with the central sunspot region (shown by the yellow lines in Figure \ref{Fig_Intro}(e)).

\begin{deluxetable*}{p{0.3cm}p{3cm}p{1cm}p{1.5cm}p{8.5cm}}
\tablenum{1}
\tablecaption{Summary of the pre- to postflare evolution of AR NOAA 11302 on 2011 September 26\label{table1}}
\tablehead{
\colhead{Sr. No.} & \colhead{Phase} & \colhead{Flux Peaks} & \colhead{Time (UT)} & \colhead{Remarks}
}
\startdata
\vspace*{-0.5 cm} 1 &
\vspace*{-1 cm}
\begin{tabular}{c}
Phase 1 \\
(04:04 UT -- 04:39 UT) \\
\end{tabular}
& 
\begin{tabular}{c}
\ P1$_1$ \\
\ P1$_2$ \\
\ P1$_3$ \\
\ P1$_4$ \\
\ P1$_5$ \\
\ P1$_6$ \\
\end{tabular} &
\begin{tabular}{c}
\ $\approx$04:08\\
\ $\approx$04:17\\
\ $\approx$04:21\\
\ $\approx$04:24\\
\ $\approx$04:32\\
\ $\approx$04:36\\
\end{tabular}&
\vspace*{-1 cm} Brightening in compact loop system associated with coronal null point topology. RHESSI missed observation of P1$_1$ due to `RHESSI night'. Localized RHESSI sources within the energy range 3--25 keV from at the location during subsequent SXR peaks. \\
\hline
\vspace*{-0.3 cm} 2 &
\vspace*{-1 cm}
\begin{tabular}{c}
Phase 2 \\
(04:39 UT -- 04:53 UT) \\
\end{tabular}
& 
\begin{tabular}{c}
\ P2$_1$ \\
\ P2$_2$ \\
\ P2$_3$ \\
\ P2$_4$ \\
\end{tabular} &
\begin{tabular}{c}
\ $\approx$04:41\\
\ $\approx$04:44\\
\ $\approx$04:49\\
\ $\approx$04:52\\
\end{tabular}&
\vspace*{-0.7 cm} Highly compact X-ray sources up to $\approx$25 keV energy from the location of null point. Eruption of small loop-like structures from the null point configuration during P2$_1$ and P2$_2$. \\
\hline
\vspace*{0.2 cm} 3 &
\vspace*{-0.4 cm}
\begin{tabular}{c}
Impulsive \\
(05:06 UT -- 05:08 UT) \\
\end{tabular}
&&& Activation and failed eruption of one filament. Development of intertwined double-decker flux rope system. X-ray sources up to energy $\approx$50 keV from the location of the activated filaments.\\
\hline
\vspace*{0.2 cm} 4 &
\vspace*{-0.4 cm}
\begin{tabular}{c}
Gradual \\
(05:08 UT -- 05:13 UT) \\
\end{tabular}
&&& Brightening of a highly structured loop system. Intense EUV emission from flare ribbons and post-reconnection loop arcade. Multiple ribbon-like brightening from regions situated north of the core flaring location.\\
\enddata
\end{deluxetable*}

\section{Multi-wavelegth Imaging of Coronal Energy Release: Onset and Consequences} \label{Sec_Results}      

In Figure \ref{Fig_GOES_AIA}, we present an overview of the prominent episodes of energy release in multiple EUV channels of AIA (171, 131, and 304 \AA ) with respect to X-ray flux evolution obtained from GOES 0.5--4 \AA\ channel and RHESSI 6-12 keV energy band. AIA images during the P1$_1$ peak (see Figure \ref{Fig_Lightcurve}) suggests a localized brightening (indicated by the red arrow in Figure \ref{Fig_GOES_AIA}(c)) situated at the northern end of the filament (indicated by the white arrows in Figure \ref{Fig_GOES_AIA}(b)--(d)). Notably, this brightening was most prominent in the high temperature AIA 131 \AA\ channel compared to the AIA filters that sample plasma at lower temperatures, e.g., 171 and 304 \AA . Emission from this localized region significantly increased in all the AIA EUV channels during the P2$_1$ and P2$_2$ peaks (indicated by the red arrows in Figures \ref{Fig_GOES_AIA}(e)--(f) and the sky-colored arrow in Figure \ref{Fig_GOES_AIA}(g)). This localized region was identified as a coronal null point configuration in the NLFFF extrapolation results (Figure \ref{Fig_Intro}), suggesting null point reconnection to be responsible for the repetitive flux enhancements during the extended periods of Phase 1 and Phase 2. During this time, we observed signatures of a second filament (indicated by the green arrows in Figure \ref{Fig_GOES_AIA}(e), (g)) close to the filament previously observed (indicated by white arrows in Figure \ref{Fig_GOES_AIA}(e), (g) and also in Figures \ref{Fig_GOES_AIA}(b)--(d)). The onset of the impulsive phase of the M4.0 flare was characterized by the activation of the first filament which resulted in intensified emission (the filament appearance changed from absorption to emission, indicated by the white arrows in Figures \ref{Fig_GOES_AIA}(h) and (j)). During the impulsive phase, the filament initially displayed eruptive motion and extended spatially along its length. However, the eruption of the filament ceased to continue within $\approx$2 min after which we observed a complex structure where both the activated filaments became intertwined with each other during the peak phase of the flare. In Figures \ref{Fig_GOES_AIA}(k) and (m), we indicate the two activated intertwined filaments by the white and green arrows. Notably, during this time we observed quite intense emission from a second location also, which was situated remotely to the north of the filaments (indicated by the sky-colored arrows in Figures \ref{Fig_GOES_AIA}(k) and (m)). A summary of the different evolutionary phases under study is provided in Table \ref{table1}.

\subsection{Episoidic Energy Release at Null Point Topology} \label{Sec_Preflare}

\begin{figure}
\epsscale{1.05}
\plotone{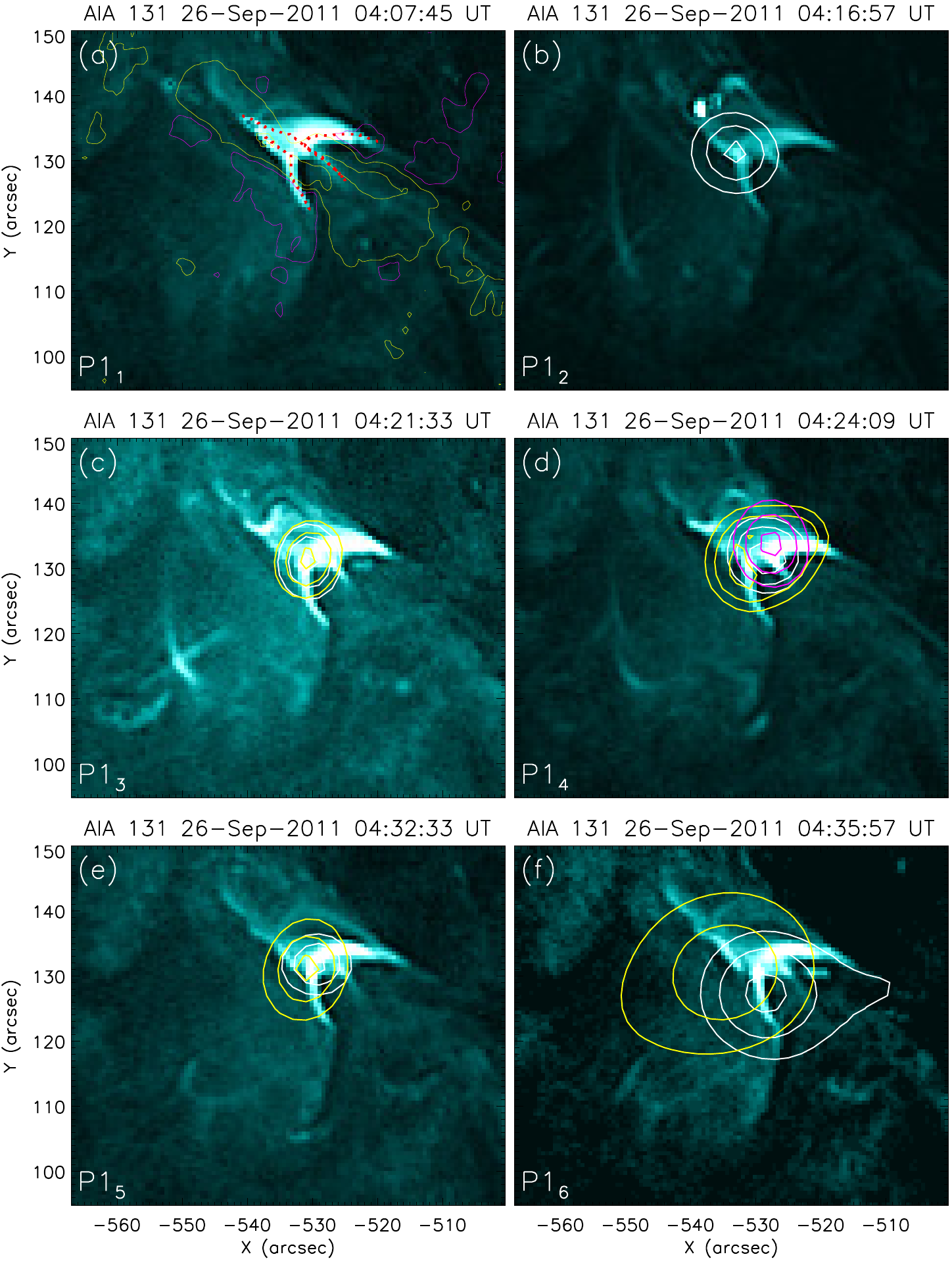}
\caption{Evolution of Phase 1 in the AIA 131 \AA\ channel. The red dotted curves in panel (a) indicate the coronal loops involved in the null point topology, during the different flux peaks of Phase 1 (i.e., P1$_1$--P1$_6$). Contours of a co-temporal HMI LOS Magnetograms are overplotted in panel (a). Contour levels are $\pm$[400, 1400] G. Yellow and purple contours refer to positive and negative field, respectively. Co-temporal RHESSI contours of 3--6 keV (white), 6--12 keV (yellow), and 12--25 keV (pink) energy bands are plotted over panels (b)--(f). Contour levels are 55\%, 75\%, and 95\% of the corresponding peak flux.}
\label{Fig_P1_131}
\end{figure}

\begin{figure}
\epsscale{1.05}
\plotone{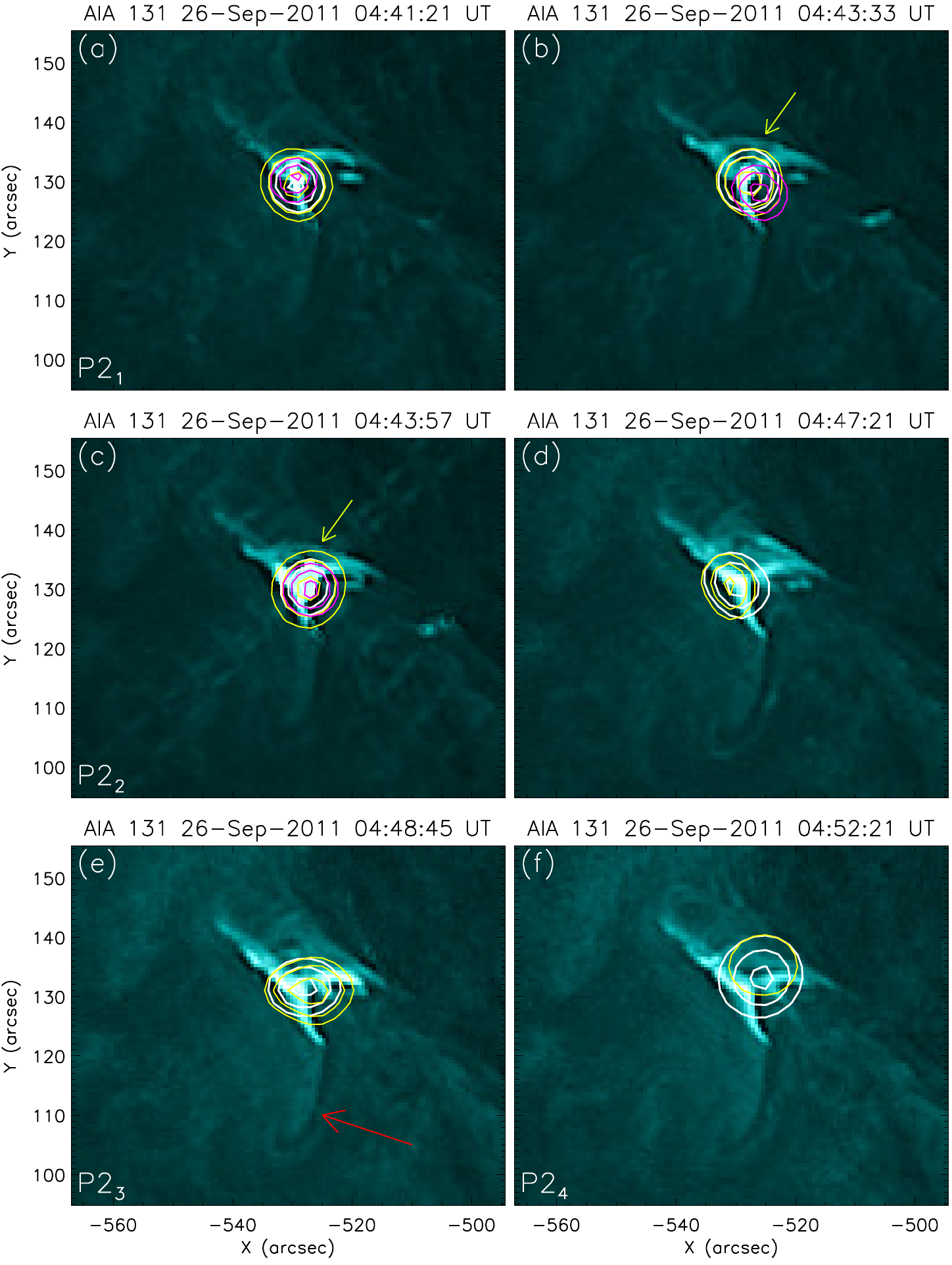}
\caption{Evolution of Phase 2 in AIA 131 \AA\ channel. Panels (a), (c), (e), and (f) are during P2$_1$--P2$_4$ peaks, respectively. Contours of co-temporal RHESSI sources in 3--6 keV (white), 6--12 keV (yellow) and 12--25 keV (pink) energy bands are plotted in selected panels. Contour levels are 55\%, 75\%, and 95\% of the corresponding peak flux. Yellow arrows in panels (b) and (c) indicate small-loop like erupting structures. The red arrow in panel (e) indicate the filament which underwent activation during the M-class flare.}
\label{Fig_P2_131}
\end{figure}

Comparison of Figures \ref{Fig_Lightcurve}(a) and (b) readily suggests that the small-scale flux enhancements observed in X-ray channels were more in agreement with hot AIA channel intensity variations (i.e., 94 and 131 \AA ) compared to the low temperature channels (i.e., 304 and 171 \AA ). Therefore, in order to have a thorough understanding of the activities during Phase 1 and Phase 2, we examine the AIA 131 \AA\ images of the flaring region in Figures \ref{Fig_P1_131} and \ref{Fig_P2_131}, respectively. Hot AIA EUV imaging clearly revealed that repetitive X-ray/EUV peaks during Phases 1 and 2 were essentially linked with an episodic, impulsive brightening of the compact loop structure. Notably, as discussed in Section \ref{Sec_Preflare_conf} (see also Figure \ref{Fig_Intro}), the region of these compact bright loops has been identified with the location of coronal null point topology.

In Figure \ref{Fig_P1_131}(a), we show AIA 131 \AA\ image of the region where we outline the bright loops by red dashed curves. From the overplotted LOS magnetogram contours (Figure \ref{Fig_P1_131}(a)), it becomes evident that the bright loops were connecting different opposite polarity regions of the central sunspot group (see also Figures \ref{Fig_Intro}(b), d)). During the subsequent emission peaks of Phase 1 i.e., P1$_2$--P1$_6$, we observed clear X-ray sources up to 25 keV which originated from the location of the compact loop system associated with the null point. Despite continued X-ray emission and enhanced loop brightening in EUV, there was no significant change in the morphology of the compact loop system. This brightness of compact EUV loop system increased during Phase 2 (Figure \ref{Fig_P2_131}) which is also evident from the EUV light curves (Figure \ref{Fig_Lightcurve}(b)). Notably, RHESSI sources were identified to be much compact during Phase 2 than in Phase 1. Further, during Phase 2, we found continued X-ray emission at higher energy (12--25 keV). Thus, the morphology and intensity of X-ray and EUV sources during Phase 2 suggest the reconnection activities at the null point to be more energetic. Also, we observed eruptions of small loop-like structures from the location of the null point during the peaks P2$_1$ and P2$_2$ (indicated by the yellow arrows in Figures \ref{Fig_P2_131}(b) and (c)) which point toward the restructuring of the magnetic configuration in the vicinity of the null point as a result of null point reconnection.

\subsection{Failed Eruption of the Filament and Associated M4.0 Flare} \label{Sec_Flare}      
After a prolonged period of repetitive events of small-scale coronal energy release,
the main M4.0 flare initiated at $\approx$05:06 UT. In Figure \ref{Fig_Flare_94}, we show a series of AIA 94 \AA\ images of the core region displaying different phases of the main flare. The pre-flare filament was not prominently visible (indicated by the red arrow in Figure \ref{Fig_Flare_94}(a)) in the AIA 94 \AA\ channel images which samples high-temperature coronal plasma ($\approx$6 MK) but during the onset of the highly impulsive rise phase of the flare, we observe narrow, stretched bright emission beneath the filament along its entire length (indicated by the black arrow in Figure \ref{Fig_Flare_94}(b)). During the impulsive phase, the activated filament went through an eruptive motion toward the projected eastern direction (see arrows in Figures \ref{Fig_Flare_94}(b)--(e)). We observed strong X-ray sources up to energy $\approx$50 keV from the location of the activated filament. Notably, the X-ray sources seem to coincide well with the initial location of the spatially elongated filament, which most likely generated from the newly formed post reconnection arcade (Figure \ref{Fig_Flare_94}(d)). During the gradual phase, the erupting filament interacted with a set of closed low-coronal loops lying above the core region which most likely ceased the eruption of the filament leading the flare in the `failed eruptive' category. The gradual phase of the flare was characterized by intense emission from those closed low-coronal loops (indicated by the blue arrow in Figure \ref{Fig_Flare_94}(f)) as well as from the post-flare arcade following the erupting filament (indicated by the black arrow in Figure \ref{Fig_Flare_94}(f)). A comparison of the AIA 94 \AA\ image with the overplotted contours of co-temporal HMI LOS magnetogram (Figure \ref{Fig_Flare_94}(f)) reveals that the overlying coronal loops connected the negative polarity regions of the central sunspot region to the dispersed positive polarity region situated to the north of the central sunspot group.

\begin{figure}
\epsscale{0.75}
\plotone{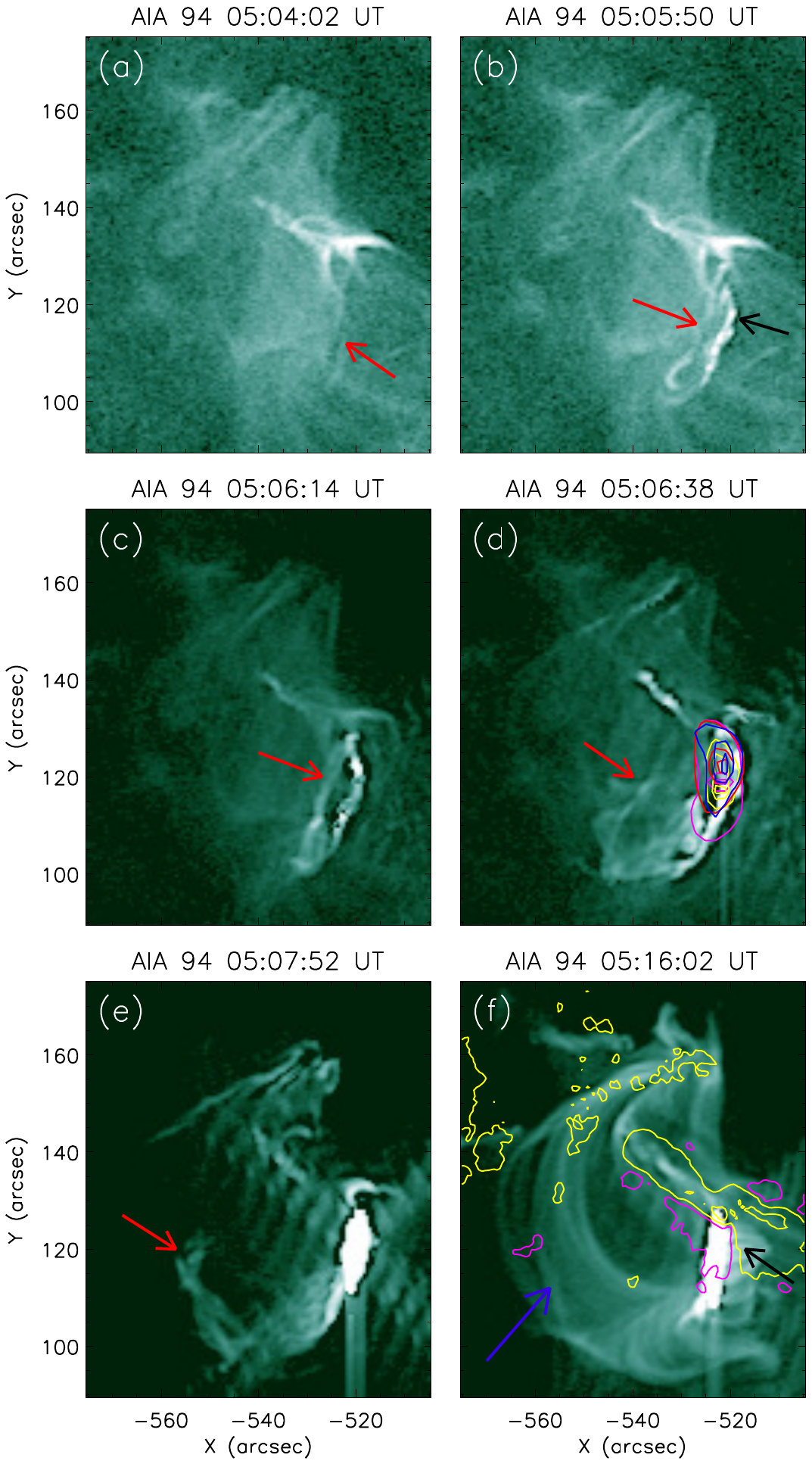}
\caption{Series of AIA 94 \AA\ images depicting the evolution of the M-class flare. The black arrow in panel (b) indicate the localized brightening suggesting the onset of the flare. The red arrows in panels (a)--(e) indicate the filament during different phases of its failed eruption. The black arrow in panel (f) indicate emission from the flare arcade. The blue arrow in panel (f) indicate highly structured coronal loops overlying the post-reconnection arcade. Contours of HMI LOS magnetogram are overplotted in panel (f). Contour levels are $\pm$(500, 1500) G. Yellow and pink contours refer to positive and negative flux, respectively. Contours of co-temporal RHESSI sources in 3--6 keV (pink), 6--12 keV (yellow), 12--25 keV (red), and 25--50 keV (blue) energy bands are plotted in selected panels. Contour levels are 50\%, 80\%, and 95\% of the corresponding peak flux.}
\label{Fig_Flare_94}
\end{figure}

In order to investigate the evolution in the low atmospheric layers of the Sun, we look at AIA 304 \AA\ images (Figure \ref{Fig_Flare_304}), where we readily observe signatures of two quite prominent filaments which appeared to be separated spatially prior to the onset of the flare (indicated by the yellow arrows in Figure \ref{Fig_Flare_304}(a)). During the impulsive phase, as the filament displayed a brief period of eruption, we observed intense emission from the core of the AR (Figures \ref{Fig_Flare_304}(b)--(c)). During the peak phase of the flare, the two filaments appeared to get intertwined with each other, resembling a double-decker flux rope system. The two hot filaments of the double-decker system are indicated by the green and sky-colored arrows in Figure \ref{Fig_Flare_304}(d). The double-decker structure reduced in spatial extension during the early gradual phase of the flare (indicated by the yellow arrows in Figure \ref{Fig_Flare_304}(e)), which was further followed by the formation of flare ribbons and post-flare arcades (within the green dashed box in Figure \ref{Fig_Flare_304}(f)).

\begin{figure}
\epsscale{0.75}
\plotone{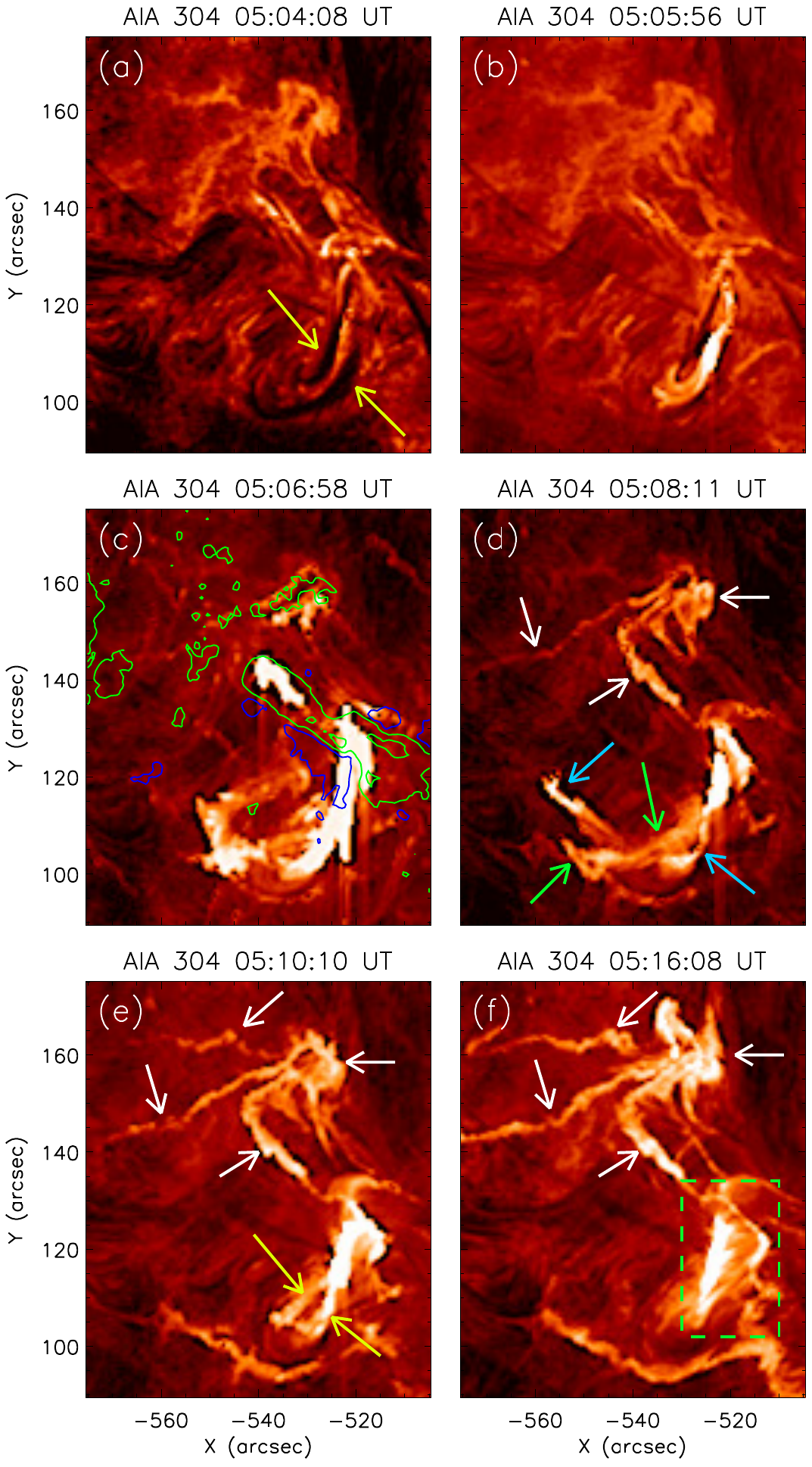}
\caption{Series of AIA 304 \AA\ images depicting the evolution of the M-class flare. The yellow arrows in panel (a) indicate the two filaments prior to the onset of the flare. The sky-blue and green arrows in panel (d) indicate the two filaments during the peak phase of the flare. The yellow arrows in panel (e) indicate the two filaments after the failed eruption. The green dashed box in panel (f) outline the core region during the gradual phase of the flare where intense emission from the flare ribbons and post-reconnection arcade can be identified. The white arrows in panels (d)--(f) indicate ribbon-like brightenings observed during the gradual phase of the flare. Contours of HMI LOS magnetogram are overplotted in panel (c). Contour levels are $\pm$(500, 1500) G. Green and blue contours refer to positive and negative flux, respectively.}
\label{Fig_Flare_304}
\end{figure}

Interestingly, during the peak phase of the M4.0 flare ($\approx$05:08 UT), we observed multiple ribbon-like brightening from regions situated to the north of location of the filaments (indicated by the white arrows in Figure \ref{Fig_Flare_304}(d)). During the gradual phase, emission from the ribbon-like structures significantly intensified (cf., Figures \ref{Fig_Flare_304}(d), (e), (f)). Topological analysis have been carried out to explore the physical connections of these remote brightenings with the flaring processes (Section \ref{Sec_NLFFF_null}).

\section{Coronal Magnetic Field Modeling} \label{Sec_Extrapolation}

\subsection{Pre-flare Coronal Configuration and Post-flare arcade} \label{Sec_NLFFF_null}

\begin{figure}
\plotone{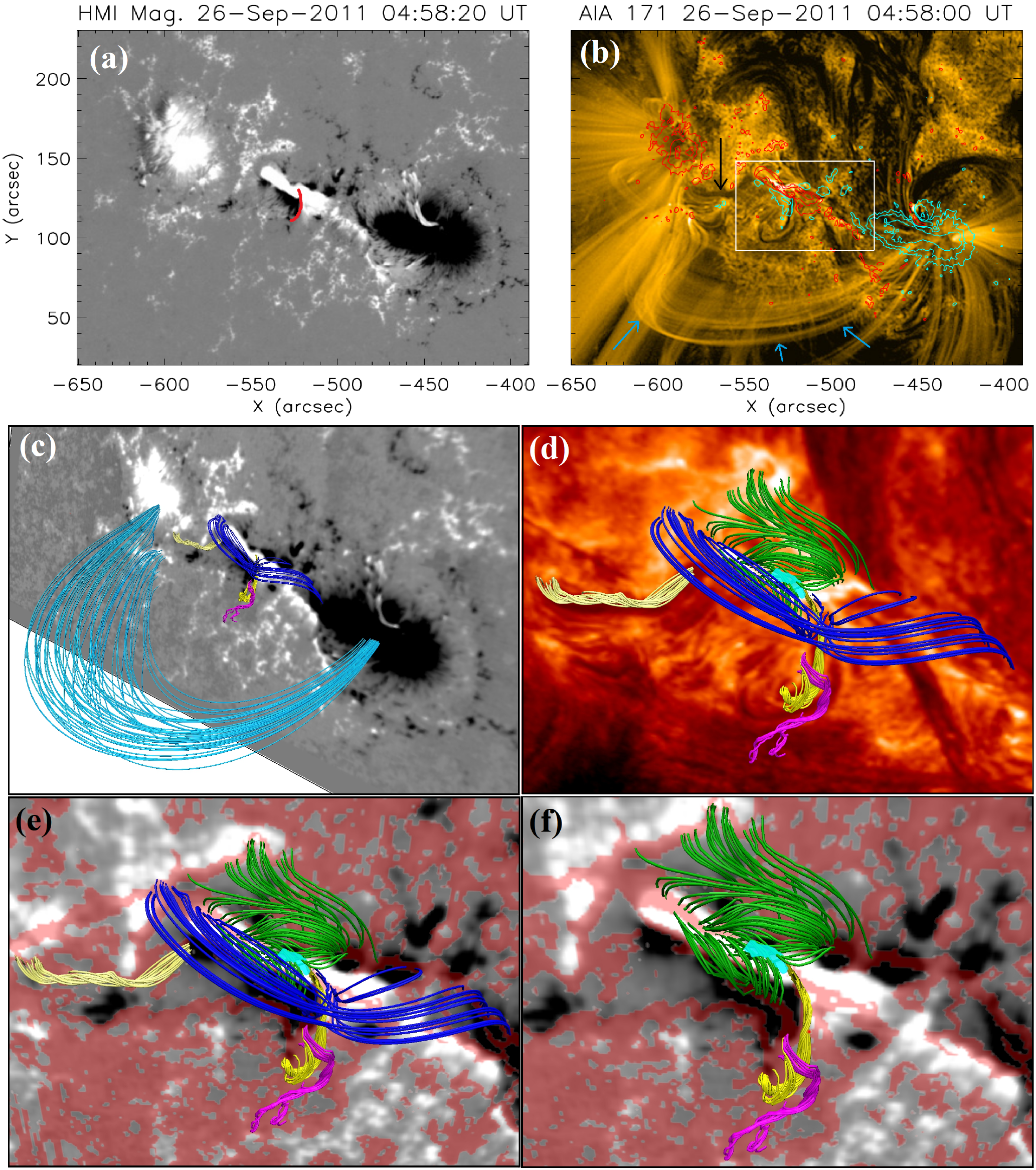}
\caption{Panel (a): HMI LOS magnetogram of AR NOAA 11302. Panel (b): Co-temporal AIA 171 \AA\ image. The flaring region is outlined by the white box. The sky-blue colored arrows indicate a set of large, high-lying coronal loops above the flaring region. Contours of the LOS magnetograms are plotted over the AIA 171 \AA\ image at levels of $\pm$[500, 1000, 1500] G. Red and cyan contours refer to positive and negative polarities, respectively. Panel (c): NLFFF extrapolated magnetic field lines showing the important coronal connectivities in the overall active region. In panels (d)--(f), we show only the model field lines in the flaring region. The bright yellow and pink lines represent two flux ropes. The northern part of the flux rope shown by bright yellow, was enveloped by the blue lines. A set of field lines (shown by green color) situated to the north of the flux rope shown in bright yellow, was associated with a high squashing factor. The sky-colored patch in panels (d)--(f) within the region of green lines is characterized by $\log(Q)=3$. The light yellow lines in panels (d) and (e) represent another flux rope which was present in the active region but did not take part in the flaring activity. In panel (f), we show the same model coronal loops except the blue and light yellow lines for better visualization of the green lines. The color template used in the background magnetograms is same as in Figure \ref{Fig_Intro}(b). Background of panel (d) is AIA 304 \AA\ image. The reddish colored patches over the background in panels (e)--(f) is characterized by $\log(Q)=2$. The red curve in panel (a) approximately denotes the axis of the bright-yellow flux rope, which is used for the magnetic decay index analysis shown in Figure \ref{Fig_Decay_saddle}.}
\label{Fig_Extrapolation_general}
\end{figure}

In Figure \ref{Fig_Extrapolation_general}, we plot NLFFF-extrapolated coronal field lines in the AR prior to the onset of the M4.0 flare and compare them with the coronal loops observed in EUV images. For convenience, we plot co-temporal LOS magnetogram and AIA 171 \AA\ image of the AR in Figures \ref{Fig_Extrapolation_general}(a) and (b), respectively. From the contours of the HMI LOS magnetogram plotted on top of AIA 171 \AA\ image (Figure \ref{Fig_Extrapolation_general}(b)), it becomes clear that, apart from the flaring region, the AR was characterized by a set of large coronal loops extended upto high coronal heights which is indicated by the sky-colored arrows. We further note the presence of a different filament within the AR which was not involved in the reported flaring activity. In Figure \ref{Fig_Extrapolation_general}(b), we have indicated the filament by a black arrow.

In Figure \ref{Fig_Extrapolation_general}(c), we display few sets of modeled coronal field lines in the overall AR. We find that the sky-colored field lines correlate well with the coronal loops indicated by the sky-colored arrows in Figure \ref{Fig_Extrapolation_general}(b). In Figures \ref{Fig_Extrapolation_general}(d)--(f), we focus to the NLFFF-extrapolated field lines within the flaring region only, where we readily identify three flux ropes which are shown in bright yellow, pink and light yellow colors. Interestingly, the flux rope manifested by the light yellow color is precisely co-spatial with the filament indicated by the black arrow in Figure \ref{Fig_Extrapolation_general}(b) which is further demonstrated by the AIA 304 \AA\ image in the background of Figure \ref{Fig_Extrapolation_general}(d). Further, the two flux ropes shown in bright yellow and pink colors constitute a double-decker flux rope configuration which is exactly co-spatial with the two filaments which were involved in the M4.0 flare (Figures \ref{Fig_GOES_AIA}, \ref{Fig_Flare_304}).

Here we recall that, during the gradual phase of the flare, multiple ribbon-like brightenings situated to the north of the core region became very bright (indicated by the white arrow in Figure \ref{Fig_Flare_304}(f); Section \ref{Sec_Flare}). NLFFF-extrapolation results suggest a set of field lines connecting that region to the northern leg of the double-decker flux rope system, which are shown in green color (Figure \ref{Fig_Extrapolation_general}(d)--(f)). Notably, few of the green lines, exactly above the northern leg of the flux rope shown in bright yellow, was characterized by high values of the squashing factor ($\log(Q)=3$) which is represented by the sky-colored patch in Figures \ref{Fig_Extrapolation_general}(d)--(f). Further, the footpoints of the green colored model coronal loops were characterized by high $Q$ values which is evident from the background red colored patches representing $\log(Q)=2$, in Figures \ref{Fig_Extrapolation_general}(e)--(f). Notably, the coronal loops involved in the quadrupolar configuration are shown by the blue colored lines in Figures \ref{Fig_Extrapolation_general}(c)--(e).

In Figure \ref{Fig_Extrapolation_null}, we show the association of the double-decker flux ropes and the quadrupolar coronal configuration from different viewing angles. From the top and side views of the whole configuration, we readily understand that the quadrupolar configuration including the null point (indicated by the green, orange, and black arrows in Figures \ref{Fig_Extrapolation_null}(a)--(c), respectively) was lying over the yellow colored flux rope only. NLFFF modeled field lines, complemented by AIA observations, clearly demonstrate that the null point is situated above the central part of the yellow flux rope (see the green and red arrows indicating the null point and the northern leg of yellow flux rope, respectively, in Figure \ref{Fig_Extrapolation_null}(a)). Further, we could identify a different set of model field lines (shown by the sky-blue color in Figure \ref{Fig_Extrapolation_null}(d)) which manifest a structure similar to the dense coronal arcade (indicated by the blue arrow in Figure \ref{Fig_Flare_94}(f)) that developed during the gradual phase of the flare. During the impulsive phase of the flare, the flux rope (shown in bright yellow) might have undergone eruptive motion towards east and interacted with the sly-blue colored lines, giving rise to the enhanced emission from the highly structured sky-blue lines.

\subsection{Decay Index and Twist Number} \label{Sec_NLFFF_twist}
EUV observations revealed that the eruption of the filament was constrained within a short while after initiation, leading to a failed eruption. In order to investigate the coronal conditions responsible for the failed eruption, we calculated the magnetic decay index (n) within the extrapolation volume.

Theoretically, if a current ring of major radius $R$ is embedded in an external magnetic field, then the ring experiences a radially outward `hoop force', because of its curvature. In stable condition, this hoop force is balanced by the inwardly directed Lorentz force. If the Lorentz force due to the external field decreases faster with $R$ than the hoop force, the system becomes unstable due to the torus instability \citep{Bateman1978}. The decay rate of the external magnetic field is quantified by the magnetic decay index ($n$). Considering the fact that the toroidal component (directed along the axis of the flux rope) of the external magnetic field does not contribute to the strapping force \citep[a nice visual depiction is provided in `Extended Data Figure 2' in][]{Myers2015}, in the ideal current-wire approach, the magnetic decay index is calculated as

\begin{equation} \label{Eq_Decay_poloidal}
n=-\frac{d \log(B_p)}{d \log(h)}
\end{equation}

\noindent where $B_p$ is the external poloidal field (along the transverse direction of the flux rope axis) and $h$ is the height.

However, in reality, flux ropes are observed to significantly differ from the simplified shape of a semi-circular current-wire \citep[see e.g.,][]{Demoulin2010, Fan2010, Olmedo2010}. Therefore, in order to calculate the magnetic decay index, we manually determined the approximate 2D projection of the flux rope axis on the photosphere (shown by the red curve in Figure \ref{Fig_Extrapolation_general}(a)). The 3D magnetic field vector at each pixel on the vertical surface above the axis were then decomposed into two components: along the direction of the axis \textit{i.e.,} the toroidal component; and perpendicular to the axis \textit{i.e.,} the poloidal component. This perpendicular component was used in Equation \ref{Eq_Decay_poloidal} for the computation of the magnetic decay index.

\begin{figure}
\epsscale{1.15}
\plotone{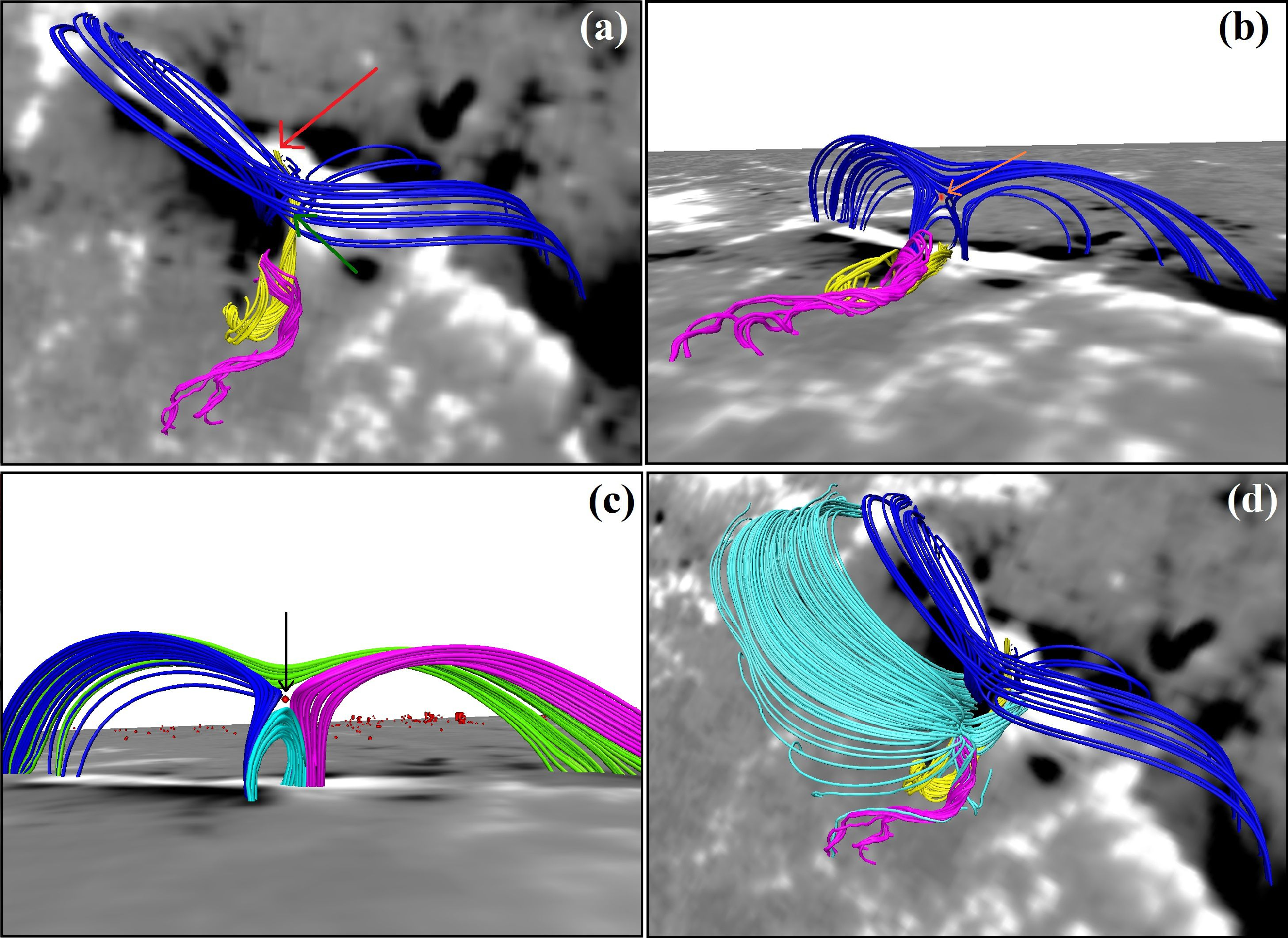}
\caption{Modeled coronal configuration involved in the pre-flare activities and the M-class flare including the two flux ropes (shown in bright yellow and pink color) and the low-coronal closed loops connecting the multi-polar photospheric configuration (shown in blue color). The northern leg of the yellow-colored flux rope is indicated by the red arrows in panels (a). The red-colored patch within the blue lines (also indicated by the orange arrow in panel (b)) is characterized by $\log(Q)=8.5$ which represents a coronal null point. The green arrow in panel (a) indicate the approximate location of the null point. In panel (c), only the representative magnetic lines involved in the null point configuration are shown. For a better representation, the four domains of magnetic field lines are shown in different colors. The black arrow indicates the null point. The sky-colored lines in panel (d) represent the set of coronal loops that intensified during the gradual phase of the flare (indicated by the blue arrow in Figure \ref{Fig_Flare_94}(f)). The color template used in the background magnetograms is same as in Figure \ref{Fig_Intro}(b).}
\label{Fig_Extrapolation_null}
\end{figure}

From the distribution of the magnetic decay index above the flux rope axis (Figure \ref{Fig_Decay_saddle}), we find that within a very low height above the flux rope, we found an extended region of high magnetic decay index (depicted by the red colored patch in Figure \ref{Fig_Decay_saddle}(a)). This region was immediately enveloped by another region where the decay index was as low as $\approx$--3 (the region shown in blue in Figure \ref{Fig_Decay_saddle}(a)). Above this region, the value of the decay index slowly increased and reached to 1.5 again at a quite high coronal layer (indicated by the yellow curve in the top portion of Figure \ref{Fig_Decay_saddle}(a)). In Figure \ref{Fig_Decay_saddle}(b), we plot the variation of the decay index averaged over the axis of the double-decker filament, with height, where we find that the average value of decay index in the high corona reached the values of 1.0 and 1.5 at heights of $\approx$43 Mm and $\approx$64 Mm, respectively.

In order to investigate the applicability of kink instability as the triggering mechanism of the flux ropes, we calculated twist number within the extrapolation volume using the IDL-based code developed by \citet{Liu2016}. The twist number \citep[$T_w$;][]{Berger2006} associated with a flux rope  is defined as

\begin{equation}
\tau_w=\int_{L}^{}\frac{(\nabla\times\vec{B})\cdot\vec{B}}{4\pi B^2} dl~,
\end{equation}

\noindent where $L$ denotes the length of the flux rope. Our calculations suggest that both the flux ropes in the double-decker flux rope system were associated with a positive twist. The average value of twist number associated with the flux rope shown in bright yellow color was found to be $\approx$1.5 while the twist number associated with the flux rope shown in pink was found to be $\approx$1.7. A statistical survey conducted by \citet{Duan2019} revealed the critical value of $|\tau_w|$ for kink instability to be 2 which suggests kink instability was not responsible for the activation of the filament during the onset of the M4.0 flares.

\begin{figure}
\epsscale{1.0}
\plotone{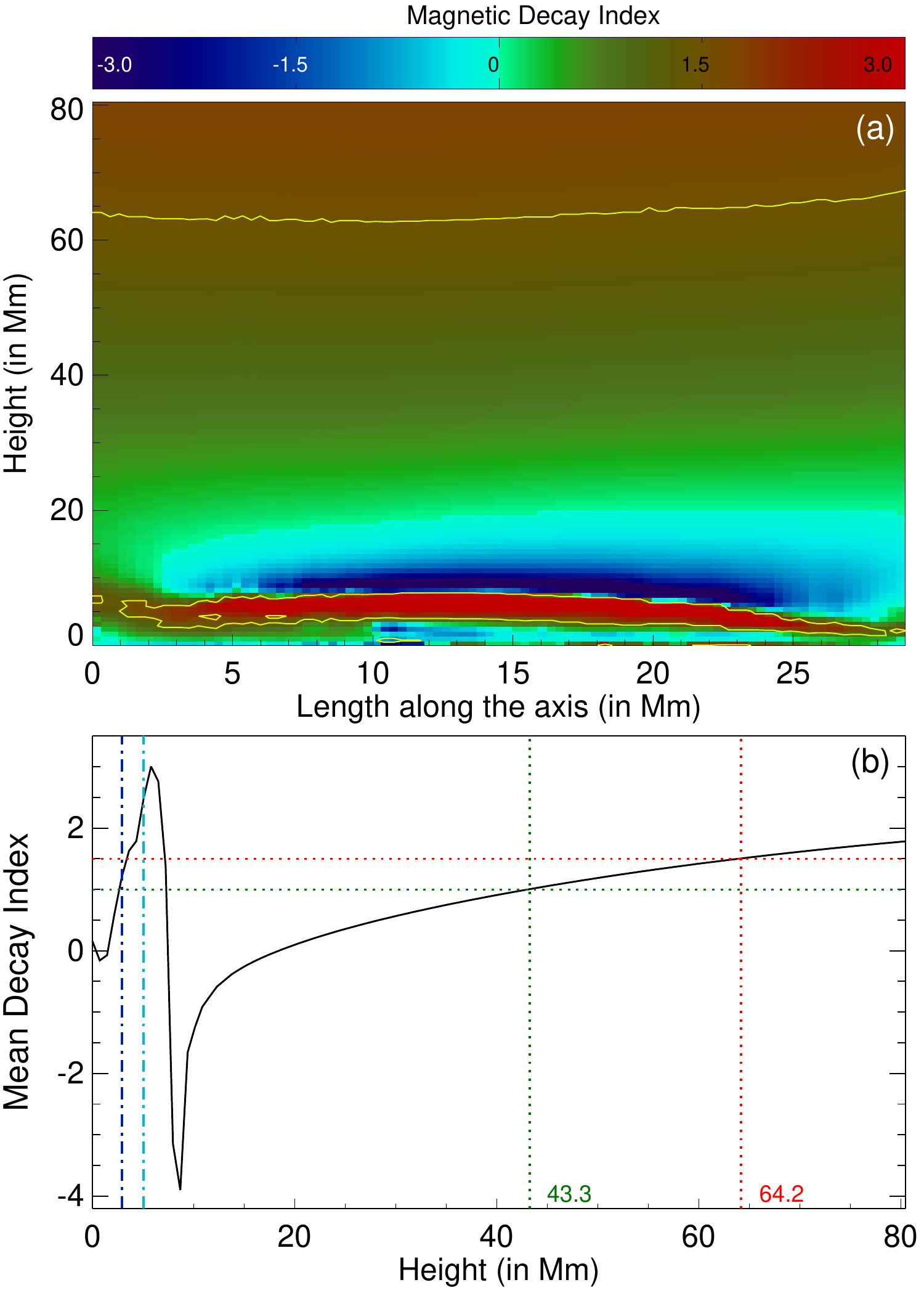}
\caption{Panel (a): Distribution of magnetic decay index with height along the vertical surface above the axis of the yellow-colored flux rope (indicated by the red curve in Figure \ref{Fig_Extrapolation_general}(a)). The yellow curves refer to $n$=1.5. The variation of decay index with height, averaged over the flux rope axis, is shown in panel (b). The green and red dotted lines indicate the critical heights corresponding to $n$=1.0 and $n$=1.5, respectively. The blue and sky-colored dash-dotted lines indicate the maximum heights of the yellow and pink-colored flux ropes, respectively (see Figure \ref{Fig_Extrapolation_null}).}
\label{Fig_Decay_saddle}
\end{figure}

\section{Discussion} \label{Sec_Discussion}
In this paper, we have investigated the triggering and subsequent failed eruption of a small filament from AR NOAA 11302. The morphology of the AR was highly interesting as three prominent and distinct sunspot groups in this AR were distributed almost in a linear manner. While the central sunspot group was the smallest among the three, it was the only bipolar sunspot group containing $\delta$-spots. Notably, the reported flare occurred from the central sunspot group.

While one filament underwent a failed eruption, EUV images of the flaring region prior to the onset of the M4.0 flare clearly revealed signatures of two distinct filaments (Figure \ref{Fig_GOES_AIA}, \ref{Fig_Flare_304}(a)). During the impulsive phase of the flare, both filaments got activated, extended spatially and developed into a complex structure where both the filaments got intertwined into each other (Figure \ref{Fig_GOES_AIA}(k)--(m), \ref{Fig_Flare_304}(d)). NLFFF extrapolation results suggest the presence of two flux ropes at the flaring region, in a double-decker flux rope configuration. The contemporary concept of double-decker flux rope system was first reported by \citet{Liu2012} and only a handful of articles have reported such complex structures since then \citep[e.g.,][]{Cheng2014, Kliem2014, Dhakal2018, Wang2018, Tian2018, Awasthi2019, Zheng2019, Mitra2020b, Mitra2021}. Despite the small number of reportings, double-decker flux rope systems can be mostly divided into two categories. While most of the reported double-decker flux rope systems were identified as two vertically well separated filaments, one of them undergoing eruption \citep[e.g.,][]{Liu2012, Dhakal2018, Tian2018, Awasthi2019, Zheng2019}, the second category comprises complex pre-flare sigmoidal structures involving intertwined flux ropes \citep{Cheng2014, Mitra2020b, Mitra2021}. In the first category, triggering and eruption of flux ropes are generally caused by photospheric activity i.e., shearing and/or rotating motion. However, in this case, the activation of one flux rope in the double-decker system may or may not influence the stability of the other flux rope \citep[see e.g.,][]{Liu2012}. On the other hand, in the case of intertwined flux ropes, interaction between the two flux ropes mostly cause activation and eruption of the system. The two filaments reported in this article were spatially well separated during the pre-flare phase. During the impulsive phase, they extended longitudinally and got intertwined with each other. Further, the onset of the flare in this case, followed clear pre-flare activities occurring at a separate location than the filaments, which was identified both in the SXR flux evolution and the AIA EUV images. In view of this, the double-decker system, reported in this study, seems to have characteristics of both the categories of double-decker systems discussed above. Such complex and intriguing behavior of magnetic flux ropes challenges our general understanding of solar magnetic fields and their evolution during transient activities.

The SXR flux evolution prior to the onset of the M-class flare revealed multiple episodes of flux enhancements (Figure \ref{Fig_Lightcurve}). Comparison of SXR lightcurves with EUV images of the AR suggested these pre-flare flux enhancements to be associated with a localized brightening that originated above the northern end of the filaments. The photospheric magnetic field of the region manifested a complex distribution of magnetic polarities involving $\delta$-spots (i.e., the central sunspot group; see Figure \ref{Fig_Intro}). NLFFF extrapolation revealed a multi-flux topology in the flaring region which forms a quadrupolar configuration with a coronal null point (Figures \ref{Fig_Intro}(e), \ref{Fig_Extrapolation_null}). Such a pre-flare coronal configuration is in agreement with the scenario prescribed in the breakout model of solar eruptions \citep{Antiochos1999, Karpen2012}. According to this model, the small-scale pre-flare reconnections at the null point will remove the overlying (constraining) magnetic flux by transferring it to the `side lobe' field lines (see Figure 1 in \citealp{Karpen2012}). In view of this, the observed pre-flare episodic brightening of the coronal loops associated with the null point is well consistent with the initiation process of the eruptive flux ropes invoked in the breakout model. We note intense and structured emission during the pre-flare phase (see Figure \ref{Fig_P2_131}) that continued till the earliest development of the M-class flare (Figure \ref{Fig_Flare_94}(a)). This structured emission can be readily perceived as the brightening of multiple low-lying loops of the quadrupolar configuration in and around the coronal mull-point, providing credence to our proposed scenario of pre-flare energy release by the null-point reconnection.

Notably the location of the null point above the filament by identifying the region of highest value of the degree of squashing factor ($Q$) within the extrapolation volume. Null points are essentially singular points in the corona where all the components of magnetic field vanishes \citep{Aschwanden2005}. Theoretically, a null point can be characterized by $Q\to\infty$ as it works as a separator between different topological domains of magnetic field \citep[see e.g.,][]{Longcope2002}. However, numerical techniques used to calculate $Q$-values will always return finite values, thus we can expect a coronal null point to be associated with high degree $Q$-values \citep{Pariat2012}. In a number of studies concerning coronal null points and their evolution, the coronal null points were found to be associated with a wide range of $Q$-values \citep[$\sim$10$^4$--10$^{12}$; see,][]{Yang2015, Liu2020, Qiu2020, Prasad2020}. In the present study, we observed the null point to be associated with a value of $Q \approx$10$^{8.5}$. Regions with smaller $Q$ are identified as `Quasi-separatrix layers' \citep[QSL;][]{Aulanier2006}. Here we remember that, during the gradual phase of the M4.0 flare, multiple ribbon-like brightenings were observed from locations north of the flaring region (Figure \ref{Fig_Flare_304}(d)--(f)). NLFFF extrapolation results revealed the presence of a set of coronal loops that connect the remote region with the region situated at the northern footpoint of the flux rope (shown by the green colored lines in Figures \ref{Fig_Extrapolation_general}(d)--(f)). Calculation of $Q$ has further revealed that the green lines are associated with strong $Q$-values (Figure \ref{Fig_Extrapolation_general}(d)--(f)). These findings demonstrate that the remote brightening during the peak phase of the main flare are an observational manifestations of slip-running reconnection influenced by the QSLs.

The most important aspect of this study is the investigation of the failed eruption of the flux ropes. Analysis of the magnetic decay index ($n$) suggests the presence of an extended region just above the flux ropes characterized by $n>1.5$ (Figure \ref{Fig_Decay_saddle}) suggestive of the flux ropes to be subject to torus instability. The topological configuration of the active region corona in our event presented a special scenario where a coronal null point existed in the core region and the region of torus instability resided immediately below the null point (Figure \ref{Fig_Extrapolation_null}(c)). High decay of magnetic field is expected below a null point which in turns increases the value of the decay index. However, the coronal region within the height range $\approx$7--20 Mm above the flux ropes (Figure \ref{Fig_Decay_saddle}(c)) is of particular interest in view of the negative magnetic decay index within this height range. Notably, the variation of average decay index with height presents a so called `saddle'-like profile where a local maximum of $n$ in lower height is followed by a local minimum of $n$ at relatively larger height \citep{Wang2017}. Such saddle-like profiles of the decay index produces a favorable condition for failed eruption of flux ropes since the toroidal strapping force (responsible to constraint the erupting magnetic structure) increases in height after an initial decrease \citep[see e.g.,][]{Wang2017, Filippov2020a}. However, a number of studies have reported successful eruption of flux ropes despite the coronal decay index profile being saddle-like \citep[e.g.,][]{Cheng2011, Wang2017, Filippov2020b} suggesting the involvement of other factors towards such successful eruption. Among a number of parameters such as, Lorentz force, non-potentiality of the source region etc, the value of decay index in the saddle bottom is believed to be important in determining the eruption profile of flux ropes \citep{Liu2018}. \citet{Inoue2018} investigated a successful flux rope eruption through a saddle-like decay index profile by numerical simulation and found that the feedback relation between the eruption of the flux rope and reconnection rate beneath it, enabled the flux rope to pass through the torus stable zone. The event reported here is a good example of a failed eruption due to a saddle-like decay index profile of the corona above the flux rope. The specialty of this event is the presence of the negative decay index region above the torus unstable zone within a low height ($\lesssim$20 Mm; Figures \ref{Fig_Decay_saddle}(c), (e)), which is rarely observed \citep[see][]{Filippov2020b}. The negative value of the magnetic decay index not only decelerated the eruption of the flux rope but also repelled it back downward.

In the present study, we analyze and interpret an intriguing case of the failed eruption of a `torus-unstable' complex double-decker flux rope system. We show that the continuation of a flux rope eruption against the overlying coronal fields is essentially controlled by the extended decay index profile characterizing the strength of strapping field through the larger coronal heights above the flux rope axis. Investigations on the failed eruptions of torus-unstable flux ropes has emerged as one of the most crucial topics in the contemporary solar physics because of their importance in the space-weather prediction. In this context, in addition to classical torus instability, some additional factors have been assessed to understand the cause of the failed eruption which are worth discussing \citep{Myers2015, Zhong2021}. Based on the results of laboratory experiment, \citet{Myers2015} found that the failed eruption can also occur when the guide magnetic field \textit{i.e.}, the toroidal field (the ambient field that runs along the flux rope) is strong enough to prevent the flux rope from kinking. Under these conditions, the guide field interacts with electric currents in the flux rope to produce a dynamic toroidal field tension force that halts the eruption. The study by \citet{Zhong2021} presents a data-driven magnetohydrodynamic simulation for a confined eruption. They showed a Lorentz force component, resulting from the radial magnetic field or the non-axisymmetry of the flux rope, which can essentially constrain the eruption. In the light of above studies and our work, we understand that the ultimate fate of solar eruptions is determined by a complex interplay of coronal magnetic fields involving the magnetic flux rope, core and envelope fields. To this end, we plan to carry out a series of subsequent studies on failed eruptions of torus unstable flux ropes in order to reach to further understandings on the mechanism(s) of successful/failed solar eruptions.

\acknowledgments
We would like to thank the SDO and RHESSI teams for their open data policy. SDO is NASA's mission under the Living With a Star (LWS) program. RHESSI was the sixth mission in the SMall EXplorer (SMEX) program of NASA. This work utilizes GONG data from NSO, which is operated by AURA under a cooperative agreement with NSF and with additional financial support from NOAA, NASA, and USAF. A.M.V acknowledges the Austrian Science Fund (FWF): project no. I4555. TW acknowledges the DLR-grant 50 OC 2101 and DFG-grant WI 3211/5-1. We are thankful to the referee for providing important comments which helped us improve the overall content of the paper.

\appendix

\section{Calculation of twist number}
For a smooth, non-self-intersecting curve $\mathbf{x}(s)$ parametrized by arclength s and a second such curve $\mathbf{y}(s)$ surrounding $\mathbf{x}$, jointly defining a ribbon, twist number $\tau_g$ is given by \citep[Equation (12) in][]{Berger2006}

\begin{equation}
\tau_g=\frac{1}{2\pi}\int_{\mathbf{x}}\mathbf{\hat{T}}(s)\cdot\mathbf{\hat{V}}(s)\times\frac{d{\mathbf{\hat{V}}(s)}}{ds}ds
\label{Eq_AB_1}
\end{equation}

\noindent where $\mathbf{\hat{T}}(s)$ represent unit tangent vector to $\mathbf{x}(s)$ and $\mathbf{\hat{V}}(s)$ denote unit vector normal to $\mathbf{x}(s)$. Here, any point on $\mathbf{y}(s)$ is related to $\mathbf{x}(s)$ by

\begin{equation}
\mathbf{y}(t)=\mathbf{x}(t)+\epsilon\mathbf{V}(t)
\label{Eq_AB_2}
\end{equation}

In the context of the present analysis, $\hat{\mathbf{T}}=\hat{\mathbf{B}}=\frac{\vec{B}}{|\vec{B}|}$, and $\vec{J}=\frac{1}{\mu_0}\vec{\nabla}\times\vec{B}$. A detailed calculation conducted by \citet{Liu2016} reveals

\begin{gather}
\frac{d\tau_g}{ds}\simeq\frac{\mu_0 J_{||}}{4\pi|\vec{B}|}+\frac{c}{2\pi|\vec{B}|} \\
\ \ \ \ \ \ \ \ \ \ \Rightarrow\tau_g=\int_{x}\frac{\mu_0 J_{||}}{4\pi|\vec{B}|}ds+\int_{x}\frac{c}{2\pi|\vec{B}|}ds
\label{Eq_AB_3}
\end{gather}

\noindent where $J_{||}=\frac{\vec{J}\cdot\vec{B}}{|\vec{B}|}$ and $c$ is a constant dependant on $\mathbf{\hat{V}}$ and spatial variation of $\vec{B}$. The first term in the right hand side approaches $\tau_w$ close to the axis of the flux rope i.e.,

\begin{equation}
\lim_{\epsilon \to 0} \tau_w(\epsilon)=\tau_g-\int_{x}\frac{c}{2\pi|\vec{B}|}ds
\label{Eq_AB_4}
\end{equation}

From Equation \ref{Eq_AB_4}, it becomes clear that $\tau_w$ provides an underestimation of the true twist number of flux ropes. However, calculation of $\tau_g$ requires to include the exact geometry of the flux rope. Finding the exact geometry of a flux rope from the extrapolated magnetic field is practically not possible. Considering the flux rope to have approximated uniform twist, i.e., an uniform $\alpha$-flux rope \citep{Lundquist1950}, $c\approx0$ i.e., $\tau_w\approx\tau_g$ \citep{Liu2016}.


\begin{thebibliography}{}
\expandafter\ifx\csname natexlab\endcsname\relax\def\natexlab#1{#1}\fi
\providecommand{\url}[1]{\href{#1}{#1}}
\providecommand{\dodoi}[1]{doi:~\href{http://doi.org/#1}{\nolinkurl{#1}}}
\providecommand{\doeprint}[1]{\href{http://ascl.net/#1}{\nolinkurl{http://ascl.net/#1}}}
\providecommand{\doarXiv}[1]{\href{https://arxiv.org/abs/#1}{\nolinkurl{https://arxiv.org/abs/#1}}}

\bibitem[{{Alexander} {et~al.}(2006){Alexander}, {Liu}, \&
  {Gilbert}}]{Alexander2006}
{Alexander}, D., {Liu}, R., \& {Gilbert}, H.~R. 2006, \apj, 653, 719,
  \dodoi{10.1086/508137}

\bibitem[{{Amari} {et~al.}(2018){Amari}, {Canou}, {Aly}, {Delyon}, \&
  {Alauzet}}]{Amari2018}
{Amari}, T., {Canou}, A., {Aly}, J.-J., {Delyon}, F., \& {Alauzet}, F. 2018,
  \nat, 554, 211, \dodoi{10.1038/nature24671}

\bibitem[{{Antiochos} {et~al.}(1994){Antiochos}, {Dahlburg}, \&
  {Klimchuk}}]{Antiochos1994}
{Antiochos}, S.~K., {Dahlburg}, R.~B., \& {Klimchuk}, J.~A. 1994, \apjl, 420,
  L41, \dodoi{10.1086/187158}

\bibitem[{{Antiochos} {et~al.}(1999){Antiochos}, {DeVore}, \&
  {Klimchuk}}]{Antiochos1999}
{Antiochos}, S.~K., {DeVore}, C.~R., \& {Klimchuk}, J.~A. 1999, \apj, 510, 485,
  \dodoi{10.1086/306563}

\bibitem[{{Archontis} \& {Hood}(2008)}]{Archontis2008}
{Archontis}, V., \& {Hood}, A.~W. 2008, \apjl, 674, L113,
  \dodoi{10.1086/529377}

\bibitem[{{Aschwanden}(2005)}]{Aschwanden2005}
{Aschwanden}, M.~J. 2005, {Physics of the Solar Corona. An Introduction with
  Problems and Solutions (2nd edition)}

\bibitem[{{Aulanier} {et~al.}(2013){Aulanier}, {D{\'e}moulin}, {Schrijver},
  {Janvier}, {Pariat}, \& {Schmieder}}]{Aulanier2013}
{Aulanier}, G., {D{\'e}moulin}, P., {Schrijver}, C.~J., {et~al.} 2013, \aap,
  549, A66, \dodoi{10.1051/0004-6361/201220406}

\bibitem[{{Aulanier} {et~al.}(2012){Aulanier}, {Janvier}, \&
  {Schmieder}}]{Aulanier2012}
{Aulanier}, G., {Janvier}, M., \& {Schmieder}, B. 2012, \aap, 543, A110,
  \dodoi{10.1051/0004-6361/201219311}

\bibitem[{{Aulanier} {et~al.}(2005){Aulanier}, {Pariat}, \&
  {D{\'e}moulin}}]{Aulanier2005}
{Aulanier}, G., {Pariat}, E., \& {D{\'e}moulin}, P. 2005, \aap, 444, 961,
  \dodoi{10.1051/0004-6361:20053600}

\bibitem[{{Aulanier} {et~al.}(2006){Aulanier}, {Pariat}, {D{\'e}moulin}, \&
  {DeVore}}]{Aulanier2006}
{Aulanier}, G., {Pariat}, E., {D{\'e}moulin}, P., \& {DeVore}, C.~R. 2006,
  \solphys, 238, 347, \dodoi{10.1007/s11207-006-0230-2}

\bibitem[{{Aulanier} {et~al.}(2010){Aulanier}, {T{\"o}r{\"o}k}, {D{\'e}moulin},
  \& {DeLuca}}]{Aulanier2010}
{Aulanier}, G., {T{\"o}r{\"o}k}, T., {D{\'e}moulin}, P., \& {DeLuca}, E.~E.
  2010, \apj, 708, 314, \dodoi{10.1088/0004-637X/708/1/314}

\bibitem[{{Awasthi} {et~al.}(2019){Awasthi}, {Liu}, \& {Wang}}]{Awasthi2019}
{Awasthi}, A.~K., {Liu}, R., \& {Wang}, Y. 2019, \apj, 872, 109,
  \dodoi{10.3847/1538-4357/aafdad}

\bibitem[{{Bateman}(1978)}]{Bateman1978}
{Bateman}, G. 1978, {MHD instabilities}

\bibitem[{{Baumgartner} {et~al.}(2018){Baumgartner}, {Thalmann}, \&
  {Veronig}}]{Baumgartner2018}
{Baumgartner}, C., {Thalmann}, J.~K., \& {Veronig}, A.~M. 2018, \apj, 853, 105,
  \dodoi{10.3847/1538-4357/aaa243}

\bibitem[{{Benz}(2017)}]{Benz2017}
{Benz}, A.~O. 2017, Living Reviews in Solar Physics, 14, 2,
  \dodoi{10.1007/s41116-016-0004-3}

\bibitem[{{Berger} \& {Prior}(2006)}]{Berger2006}
{Berger}, M.~A., \& {Prior}, C. 2006, Journal of Physics A Mathematical
  General, 39, 8321, \dodoi{10.1088/0305-4470/39/26/005}

\bibitem[{{Carmichael}(1964)}]{Carmichael1964}
{Carmichael}, H. 1964, NASA Special Publication, 50, 451

\bibitem[{{Chatterjee} \& {Fan}(2013)}]{Chatterjee2013}
{Chatterjee}, P., \& {Fan}, Y. 2013, \apjl, 778, L8,
  \dodoi{10.1088/2041-8205/778/1/L8}

\bibitem[{{Chen}(2011)}]{Chen2011}
{Chen}, P.~F. 2011, Living Reviews in Solar Physics, 8, 1,
  \dodoi{10.12942/lrsp-2011-1}

\bibitem[{{Cheng} {et~al.}(2014){Cheng}, {Ding}, {Zhang}, {Sun}, {Guo}, {Wang},
  {Kliem}, \& {Deng}}]{Cheng2014}
{Cheng}, X., {Ding}, M.~D., {Zhang}, J., {et~al.} 2014, \apj, 789, 93,
  \dodoi{10.1088/0004-637X/789/2/93}

\bibitem[{{Cheng} {et~al.}(2011){Cheng}, {Zhang}, {Ding}, {Guo}, \&
  {Su}}]{Cheng2011}
{Cheng}, X., {Zhang}, J., {Ding}, M.~D., {Guo}, Y., \& {Su}, J.~T. 2011, \apj,
  732, 87, \dodoi{10.1088/0004-637X/732/2/87}

\bibitem[{{Cheng} {et~al.}(2013){Cheng}, {Zhang}, {Ding}, {Liu}, \&
  {Poomvises}}]{Cheng2013}
{Cheng}, X., {Zhang}, J., {Ding}, M.~D., {Liu}, Y., \& {Poomvises}, W. 2013,
  \apj, 763, 43, \dodoi{10.1088/0004-637X/763/1/43}

\bibitem[{{Clyne} {et~al.}(2007){Clyne}, {Mininni}, {Norton}, \&
  {Rast}}]{Clyne2007}
{Clyne}, J., {Mininni}, P., {Norton}, A., \& {Rast}, M. 2007, New Journal of
  Physics, 9, 301, \dodoi{10.1088/1367-2630/9/8/301}

\bibitem[{{D{\'e}moulin} \& {Aulanier}(2010)}]{Demoulin2010}
{D{\'e}moulin}, P., \& {Aulanier}, G. 2010, \apj, 718, 1388,
  \dodoi{10.1088/0004-637X/718/2/1388}

\bibitem[{{Demoulin} {et~al.}(1996){Demoulin}, {Henoux}, {Priest}, \&
  {Mandrini}}]{Demoulin1996}
{Demoulin}, P., {Henoux}, J.~C., {Priest}, E.~R., \& {Mandrini}, C.~H. 1996,
  \aap, 308, 643

\bibitem[{{DeRosa} {et~al.}(2015){DeRosa}, {Wheatland}, {Leka}, {Barnes},
  {Amari}, {Canou}, {Gilchrist}, {Thalmann}, {Valori}, {Wiegelmann},
  {Schrijver}, {Malanushenko}, {Sun}, \& {R{\'e}gnier}}]{DeRosa2015}
{DeRosa}, M.~L., {Wheatland}, M.~S., {Leka}, K.~D., {et~al.} 2015, \apj, 811,
  107, \dodoi{10.1088/0004-637X/811/2/107}

\bibitem[{{Dhakal} {et~al.}(2018){Dhakal}, {Chintzoglou}, \&
  {Zhang}}]{Dhakal2018}
{Dhakal}, S.~K., {Chintzoglou}, G., \& {Zhang}, J. 2018, \apj, 860, 35,
  \dodoi{10.3847/1538-4357/aac028}

\bibitem[{{Duan} {et~al.}(2019){Duan}, {Jiang}, {He}, {Feng}, {Zou}, \&
  {Cui}}]{Duan2019}
{Duan}, A., {Jiang}, C., {He}, W., {et~al.} 2019, \apj, 884, 73,
  \dodoi{10.3847/1538-4357/ab3e33}

\bibitem[{{Fan}(2010)}]{Fan2010}
{Fan}, Y. 2010, \apj, 719, 728, \dodoi{10.1088/0004-637X/719/1/728}

\bibitem[{{Filippov}(2020{\natexlab{a}})}]{Filippov2020b}
{Filippov}, B. 2020{\natexlab{a}}, \mnras, 494, 2166,
  \dodoi{10.1093/mnras/staa896}

\bibitem[{{Filippov}(2020{\natexlab{b}})}]{Filippov2020a}
{Filippov}, B.~P. 2020{\natexlab{b}}, Astronomy Reports, 64, 272,
  \dodoi{10.1134/S106377292002002X}

\bibitem[{{Fletcher} {et~al.}(2011){Fletcher}, {Dennis}, {Hudson}, {Krucker},
  {Phillips}, {Veronig}, {Battaglia}, {Bone}, {Caspi}, {Chen}, {Gallagher},
  {Grigis}, {Ji}, {Liu}, {Milligan}, \& {Temmer}}]{Fletcher2011}
{Fletcher}, L., {Dennis}, B.~R., {Hudson}, H.~S., {et~al.} 2011, \ssr, 159, 19,
  \dodoi{10.1007/s11214-010-9701-8}

\bibitem[{{Gibson}(2015)}]{Gibson2015}
{Gibson}, S. 2015, {Coronal Cavities: Observations and Implications for the
  Magnetic Environment of Prominences}, ed. J.-C. {Vial} \& O.~{Engvold}, Vol.
  415, 323, \dodoi{10.1007/978-3-319-10416-4_13}

\bibitem[{{Gibson}(2018)}]{Gibson2018}
{Gibson}, S.~E. 2018, Living Reviews in Solar Physics, 15, 7,
  \dodoi{10.1007/s41116-018-0016-2}

\bibitem[{{Gibson} \& {Fan}(2006)}]{Gibson2006}
{Gibson}, S.~E., \& {Fan}, Y. 2006, Journal of Geophysical Research (Space
  Physics), 111, A12103, \dodoi{10.1029/2006JA011871}

\bibitem[{{Gilbert} {et~al.}(2007){Gilbert}, {Alexander}, \&
  {Liu}}]{Gilbert2007}
{Gilbert}, H.~R., {Alexander}, D., \& {Liu}, R. 2007, \solphys, 245, 287,
  \dodoi{10.1007/s11207-007-9045-z}

\bibitem[{{Gou} {et~al.}(2019){Gou}, {Liu}, {Kliem}, {Wang}, \&
  {Veronig}}]{Guo2019}
{Gou}, T., {Liu}, R., {Kliem}, B., {Wang}, Y., \& {Veronig}, A.~M. 2019,
  Science Advances, 5, 7004, \dodoi{10.1126/sciadv.aau7004}

\bibitem[{{Green} {et~al.}(2018){Green}, {T{\"o}r{\"o}k}, {Vr{\v{s}}nak},
  {Manchester}, \& {Veronig}}]{Green2018}
{Green}, L.~M., {T{\"o}r{\"o}k}, T., {Vr{\v{s}}nak}, B., {Manchester}, W., \&
  {Veronig}, A. 2018, \ssr, 214, 46, \dodoi{10.1007/s11214-017-0462-5}

\bibitem[{{Harvey} {et~al.}(1996){Harvey}, {Hill}, {Hubbard}, {Kennedy},
  {Leibacher}, {Pintar}, {Gilman}, {Noyes}, {Title}, {Toomre}, {Ulrich},
  {Bhatnagar}, {Kennewell}, {Marquette}, {Patron}, {Saa}, \&
  {Yasukawa}}]{Harvey1996}
{Harvey}, J.~W., {Hill}, F., {Hubbard}, R.~P., {et~al.} 1996, Science, 272,
  1284, \dodoi{10.1126/science.272.5266.1284}

\bibitem[{{Harvey} {et~al.}(2011){Harvey}, {Bolding}, {Clark}, {Hauth}, {Hill},
  {Kroll}, {Luis}, {Mills}, {Purdy}, {Henney}, {Holland}, \&
  {Winter}}]{harvey2011}
{Harvey}, J.~W., {Bolding}, J., {Clark}, R., {et~al.} 2011, in AAS/Solar
  Physics Division Abstracts \#42, AAS/Solar Physics Division Meeting, 17.45

\bibitem[{Haynes \& Parnell(2007)}]{Haynes2007}
Haynes, A.~L., \& Parnell, C.~E. 2007, Physics of Plasmas, 14, 082107,
  \dodoi{10.1063/1.2756751}

\bibitem[{{Hernandez-Perez} {et~al.}(2019){Hernandez-Perez}, {Su}, {Veronig},
  {Thalmann}, {G{\"o}m{\"o}ry}, \& {Joshi}}]{Hernandez2019}
{Hernandez-Perez}, A., {Su}, Y., {Veronig}, A.~M., {et~al.} 2019, \apj, 874,
  122, \dodoi{10.3847/1538-4357/ab09ed}

\bibitem[{Hirayama(1974)}]{Hirayama1974}
Hirayama, T. 1974, Solar Physics, 34, 323, \dodoi{10.1007/BF00153671}

\bibitem[{{Inoue} {et~al.}(2018){Inoue}, {Kusano}, {B{\"u}chner}, \&
  {Sk{\'a}la}}]{Inoue2018}
{Inoue}, S., {Kusano}, K., {B{\"u}chner}, J., \& {Sk{\'a}la}, J. 2018, Nature
  Communications, 9, 174, \dodoi{10.1038/s41467-017-02616-8}

\bibitem[{{Janvier} {et~al.}(2013){Janvier}, {Aulanier}, {Pariat}, \&
  {D{\'e}moulin}}]{Janvier2013}
{Janvier}, M., {Aulanier}, G., {Pariat}, E., \& {D{\'e}moulin}, P. 2013, \aap,
  555, A77, \dodoi{10.1051/0004-6361/201321164}

\bibitem[{{Ji} {et~al.}(2003){Ji}, {Wang}, {Schmahl}, {Moon}, \&
  {Jiang}}]{Ji2003}
{Ji}, H., {Wang}, H., {Schmahl}, E.~J., {Moon}, Y.~J., \& {Jiang}, Y. 2003,
  \apjl, 595, L135, \dodoi{10.1086/378178}

\bibitem[{{Joshi} {et~al.}(2018){Joshi}, {Ibrahim}, {Shanmugaraju}, \&
  {Chakrabarty}}]{Joshi2018}
{Joshi}, B., {Ibrahim}, M.~S., {Shanmugaraju}, A., \& {Chakrabarty}, D. 2018,
  \solphys, 293, 107, \dodoi{10.1007/s11207-018-1325-2}

\bibitem[{{Joshi} {et~al.}(2017){Joshi}, {Kushwaha}, {Veronig}, {Dhara},
  {Shanmugaraju}, \& {Moon}}]{Joshi2017}
{Joshi}, B., {Kushwaha}, U., {Veronig}, A.~M., {et~al.} 2017, \apj, 834, 42,
  \dodoi{10.3847/1538-4357/834/1/42}

\bibitem[{{Joshi} {et~al.}(2019){Joshi}, {Zhu}, {Schmieder}, {Aulanier},
  {Janvier}, {Joshi}, {Magara}, {Chandra}, \& {Inoue}}]{Joshi2019}
{Joshi}, N.~C., {Zhu}, X., {Schmieder}, B., {et~al.} 2019, \apj, 871, 165,
  \dodoi{10.3847/1538-4357/aaf3b5}

\bibitem[{{Karpen} {et~al.}(2012){Karpen}, {Antiochos}, \&
  {DeVore}}]{Karpen2012}
{Karpen}, J.~T., {Antiochos}, S.~K., \& {DeVore}, C.~R. 2012, \apj, 760, 81,
  \dodoi{10.1088/0004-637X/760/1/81}

\bibitem[{{Kharayat} {et~al.}(2021){Kharayat}, {Joshi}, {Mitra}, {Manoharan},
  \& {Monstein}}]{Kharayat2021}
{Kharayat}, H., {Joshi}, B., {Mitra}, P.~K., {Manoharan}, P.~K., \& {Monstein},
  C. 2021, \solphys, 296, 99, \dodoi{0.1007/s11207-021-01830-4}

\bibitem[{{Kliem} {et~al.}(2013){Kliem}, {Su}, {van Ballegooijen}, \&
  {DeLuca}}]{Kliem2013}
{Kliem}, B., {Su}, Y.~N., {van Ballegooijen}, A.~A., \& {DeLuca}, E.~E. 2013,
  \apj, 779, 129, \dodoi{10.1088/0004-637X/779/2/129}

\bibitem[{{Kliem} \& {T{\"o}r{\"o}k}(2006)}]{Torok2006}
{Kliem}, B., \& {T{\"o}r{\"o}k}, T. 2006, Physical Review Letters, 96, 255002,
  \dodoi{10.1103/PhysRevLett.96.255002}

\bibitem[{{Kliem} {et~al.}(2014){Kliem}, {T{\"o}r{\"o}k}, {Titov}, {Lionello},
  {Linker}, {Liu}, {Liu}, \& {Wang}}]{Kliem2014}
{Kliem}, B., {T{\"o}r{\"o}k}, T., {Titov}, V.~S., {et~al.} 2014, \apj, 792,
  107, \dodoi{10.1088/0004-637X/792/2/107}

\bibitem[{{Kopp} \& {Pneuman}(1976)}]{Kopp1976}
{Kopp}, R.~A., \& {Pneuman}, G.~W. 1976, \solphys, 50, 85,
  \dodoi{10.1007/BF00206193}

\bibitem[{{Koskinen} {et~al.}(2017){Koskinen}, {Baker}, {Balogh}, {Gombosi},
  {Veronig}, \& {von Steiger}}]{Koskinen2017}
{Koskinen}, H. E.~J., {Baker}, D.~N., {Balogh}, A., {et~al.} 2017, \ssr, 212,
  1137, \dodoi{10.1007/s11214-017-0390-4}

\bibitem[{{Kushwaha} {et~al.}(2014){Kushwaha}, {Joshi}, {Cho}, {Veronig},
  {Tiwari}, \& {Mathew}}]{Kushwaha2014}
{Kushwaha}, U., {Joshi}, B., {Cho}, K.-S., {et~al.} 2014, \apj, 791, 23,
  \dodoi{10.1088/0004-637X/791/1/23}

\bibitem[{{Kushwaha} {et~al.}(2015){Kushwaha}, {Joshi}, {Veronig}, \&
  {Moon}}]{Kushwaha2015}
{Kushwaha}, U., {Joshi}, B., {Veronig}, A.~M., \& {Moon}, Y.-J. 2015, \apj,
  807, 101, \dodoi{10.1088/0004-637X/807/1/101}

\bibitem[{{Lanzerotti}(2017)}]{Lanzerotti2017}
{Lanzerotti}, L.~J. 2017, \ssr, 212, 1253, \dodoi{10.1007/s11214-017-0408-y}

\bibitem[{{Lemen} {et~al.}(2012){Lemen}, {Title}, {Akin}, {Boerner}, {Chou},
  {Drake}, {Duncan}, {Edwards}, {Friedlaender}, {Heyman}, {Hurlburt}, {Katz},
  {Kushner}, {Levay}, {Lindgren}, {Mathur}, {McFeaters}, {Mitchell}, {Rehse},
  {Schrijver}, {Springer}, {Stern}, {Tarbell}, {Wuelser}, {Wolfson}, {Yanari},
  {Bookbinder}, {Cheimets}, {Caldwell}, {Deluca}, {Gates}, {Golub}, {Park},
  {Podgorski}, {Bush}, {Scherrer}, {Gummin}, {Smith}, {Auker}, {Jerram},
  {Pool}, {Soufli}, {Windt}, {Beardsley}, {Clapp}, {Lang}, \&
  {Waltham}}]{Lemen2012}
{Lemen}, J.~R., {Title}, A.~M., {Akin}, D.~J., {et~al.} 2012, \solphys, 275,
  17, \dodoi{10.1007/s11207-011-9776-8}

\bibitem[{{Li} {et~al.}(2020){Li}, {Hou}, {Yang}, {Zhang}, {Liu}, \&
  {Veronig}}]{Li2020}
{Li}, T., {Hou}, Y., {Yang}, S., {et~al.} 2020, \apj, 900, 128,
  \dodoi{10.3847/1538-4357/aba6ef}

\bibitem[{{Lin} {et~al.}(2002){Lin}, {Dennis}, {Hurford}, {Smith}, {Zehnder},
  {Harvey}, {Curtis}, {Pankow}, {Turin}, {Bester}, {Csillaghy}, {Lewis},
  {Madden}, {van Beek}, {Appleby}, {Raudorf}, {McTiernan}, {Ramaty}, {Schmahl},
  {Schwartz}, {Krucker}, {Abiad}, {Quinn}, {Berg}, {Hashii}, {Sterling},
  {Jackson}, {Pratt}, {Campbell}, {Malone}, {Landis}, {Barrington-Leigh},
  {Slassi-Sennou}, {Cork}, {Clark}, {Amato}, {Orwig}, {Boyle}, {Banks},
  {Shirey}, {Tolbert}, {Zarro}, {Snow}, {Thomsen}, {Henneck}, {McHedlishvili},
  {Ming}, {Fivian}, {Jordan}, {Wanner}, {Crubb}, {Preble}, {Matranga}, {Benz},
  {Hudson}, {Canfield}, {Holman}, {Crannell}, {Kosugi}, {Emslie}, {Vilmer},
  {Brown}, {Johns-Krull}, {Aschwanden}, {Metcalf}, \& {Conway}}]{Lin2002}
{Lin}, R.~P., {Dennis}, B.~R., {Hurford}, G.~J., {et~al.} 2002, \solphys, 210,
  3, \dodoi{10.1023/A:1022428818870}

\bibitem[{{Liu} {et~al.}(2020){Liu}, {Prasad}, {Lee}, \& {Wang}}]{Liu2020}
{Liu}, C., {Prasad}, A., {Lee}, J., \& {Wang}, H. 2020, \apj, 899, 34,
  \dodoi{10.3847/1538-4357/ab9cbe}

\bibitem[{{Liu} {et~al.}(2018){Liu}, {Wang}, {Zhou}, {Dissauer}, {Temmer}, \&
  {Cui}}]{Liu2018}
{Liu}, L., {Wang}, Y., {Zhou}, Z., {et~al.} 2018, \apj, 858, 121,
  \dodoi{10.3847/1538-4357/aabba2}

\bibitem[{{Liu} {et~al.}(2012){Liu}, {Kliem}, {T{\"o}r{\"o}k}, {Liu}, {Titov},
  {Lionello}, {Linker}, \& {Wang}}]{Liu2012}
{Liu}, R., {Kliem}, B., {T{\"o}r{\"o}k}, T., {et~al.} 2012, \apj, 756, 59,
  \dodoi{10.1088/0004-637X/756/1/59}

\bibitem[{{Liu} {et~al.}(2016){Liu}, {Kliem}, {Titov}, {Chen}, {Wang}, {Wang},
  {Liu}, {Xu}, \& {Wiegelmann}}]{Liu2016}
{Liu}, R., {Kliem}, B., {Titov}, V.~S., {et~al.} 2016, \apj, 818, 148,
  \dodoi{10.3847/0004-637X/818/2/148}

\bibitem[{{Liu}(2008)}]{Liu2008}
{Liu}, Y. 2008, \apjl, 679, L151, \dodoi{10.1086/589282}

\bibitem[{{Liu} {et~al.}(2009){Liu}, {Su}, {Xu}, {Lin}, {Shibata}, \&
  {Kurokawa}}]{Liu2009}
{Liu}, Y., {Su}, J., {Xu}, Z., {et~al.} 2009, \apjl, 696, L70,
  \dodoi{10.1088/0004-637X/696/1/L70}

\bibitem[{{Longcope} \& {Klapper}(2002)}]{Longcope2002}
{Longcope}, D.~W., \& {Klapper}, I. 2002, \apj, 579, 468,
  \dodoi{10.1086/342750}

\bibitem[{Lundquist(1950)}]{Lundquist1950}
Lundquist, S. 1950, Arkiv for Fysik, B2, 361.
\newblock \url{https://ci.nii.ac.jp/naid/10007639617/en/}

\bibitem[{{Manoharan} {et~al.}(1996){Manoharan}, {van Driel-Gesztelyi}, {Pick},
  \& {Demoulin}}]{Manoharan1996}
{Manoharan}, P.~K., {van Driel-Gesztelyi}, L., {Pick}, M., \& {Demoulin}, P.
  1996, \apjl, 468, L73, \dodoi{10.1086/310221}

\bibitem[{{Martin}(1998)}]{Martin1998}
{Martin}, S.~F. 1998, \solphys, 182, 107, \dodoi{10.1023/A:1005026814076}

\bibitem[{{Masuda} {et~al.}(1994){Masuda}, {Kosugi}, {Hara}, {Tsuneta}, \&
  {Ogawara}}]{Masuda1994}
{Masuda}, S., {Kosugi}, T., {Hara}, H., {Tsuneta}, S., \& {Ogawara}, Y. 1994,
  \nat, 371, 495, \dodoi{10.1038/371495a0}

\bibitem[{{Metcalf} {et~al.}(1996){Metcalf}, {Hudson}, {Kosugi}, {Puetter}, \&
  {Pina}}]{Metcalf1996}
{Metcalf}, T.~R., {Hudson}, H.~S., {Kosugi}, T., {Puetter}, R.~C., \& {Pina},
  R.~K. 1996, \apj, 466, 585, \dodoi{10.1086/177533}

\bibitem[{{Miklenic} {et~al.}(2007){Miklenic}, {Veronig}, {Vr{\v{s}}nak}, \&
  {Hanslmeier}}]{Miklenic2007}
{Miklenic}, C.~H., {Veronig}, A.~M., {Vr{\v{s}}nak}, B., \& {Hanslmeier}, A.
  2007, \aap, 461, 697, \dodoi{10.1051/0004-6361:20065751}

\bibitem[{{Mitra} \& {Joshi}(2019)}]{Mitra2019}
{Mitra}, P.~K., \& {Joshi}, B. 2019, \apj, 884, 46,
  \dodoi{10.3847/1538-4357/ab3a96}

\bibitem[{{Mitra} \& {Joshi}(2021)}]{Mitra2021}
---. 2021, \mnras, 503, 1017, \dodoi{10.1093/mnras/stab175}

\bibitem[{{Mitra} {et~al.}(2020{\natexlab{a}}){Mitra}, {Joshi}, \&
  {Prasad}}]{Mitra2020}
{Mitra}, P.~K., {Joshi}, B., \& {Prasad}, A. 2020{\natexlab{a}}, \solphys, 295,
  29, \dodoi{10.1007/s11207-020-1596-2}

\bibitem[{{Mitra} {et~al.}(2018){Mitra}, {Joshi}, {Prasad}, {Veronig}, \&
  {Bhattacharyya}}]{Mitra2018}
{Mitra}, P.~K., {Joshi}, B., {Prasad}, A., {Veronig}, A.~M., \&
  {Bhattacharyya}, R. 2018, \apj, 869, 69, \dodoi{10.3847/1538-4357/aaed26}

\bibitem[{{Mitra} {et~al.}(2020{\natexlab{b}}){Mitra}, {Joshi}, {Veronig},
  {Chand ra}, {Dissauer}, \& {Wiegelmann}}]{Mitra2020b}
{Mitra}, P.~K., {Joshi}, B., {Veronig}, A.~M., {et~al.} 2020{\natexlab{b}},
  \apj, 900, 23, \dodoi{10.3847/1538-4357/aba900}

\bibitem[{{Moldwin}(2008)}]{Moldwin2008}
{Moldwin}, M. 2008, {An Introduction to Space Weather}

\bibitem[{{Moore} \& {Roumeliotis}(1992)}]{Moore1992}
{Moore}, R.~L., \& {Roumeliotis}, G. 1992, in Lecture Notes in Physics, Berlin
  Springer Verlag, Vol. 399, IAU Colloq. 133: Eruptive Solar Flares, ed.
  Z.~{Svestka}, B.~V. {Jackson}, \& M.~E. {Machado}, 69,
  \dodoi{10.1007/3-540-55246-4_79}

\bibitem[{{Moore} {et~al.}(2001){Moore}, {Sterling}, {Hudson}, \&
  {Lemen}}]{Moore2001}
{Moore}, R.~L., {Sterling}, A.~C., {Hudson}, H.~S., \& {Lemen}, J.~R. 2001,
  \apj, 552, 833, \dodoi{10.1086/320559}

\bibitem[{{Myers} {et~al.}(2015){Myers}, {Yamada}, {Ji}, {Yoo}, {Fox},
  {Jara-Almonte}, {Savcheva}, \& {Deluca}}]{Myers2015}
{Myers}, C.~E., {Yamada}, M., {Ji}, H., {et~al.} 2015, \nat, 528, 526,
  \dodoi{10.1038/nature16188}

\bibitem[{{Nakajima} {et~al.}(1994){Nakajima}, {Nishio}, {Enome}, {Shibasaki},
  {Takano}, {Hanaoka}, {Torii}, {Sekiguchi}, {Bushimata}, {Kawashima},
  {Shinohara}, {Irimajiri}, {Koshiishi}, {Kosugi}, {Shiomi}, {Sawa}, \&
  {Kai}}]{Nakajima1994}
{Nakajima}, H., {Nishio}, M., {Enome}, S., {et~al.} 1994, IEEE Proceedings, 82,
  705

\bibitem[{{Nindos} {et~al.}(2015){Nindos}, {Patsourakos}, {Vourlidas}, \&
  {Tagikas}}]{Nindos2015}
{Nindos}, A., {Patsourakos}, S., {Vourlidas}, A., \& {Tagikas}, C. 2015, \apj,
  808, 117, \dodoi{10.1088/0004-637X/808/2/117}

\bibitem[{{Olmedo} \& {Zhang}(2010)}]{Olmedo2010}
{Olmedo}, O., \& {Zhang}, J. 2010, \apj, 718, 433,
  \dodoi{10.1088/0004-637X/718/1/433}

\bibitem[{{Pariat} \& {D{\'e}moulin}(2012)}]{Pariat2012}
{Pariat}, E., \& {D{\'e}moulin}, P. 2012, \aap, 541, A78,
  \dodoi{10.1051/0004-6361/201118515}

\bibitem[{{Patsourakos} {et~al.}(2020){Patsourakos}, {Vourlidas},
  {T{\"o}r{\"o}k}, {Kliem}, {Antiochos}, {Archontis}, {Aulanier}, {Cheng},
  {Chintzoglou}, {Georgoulis}, {Green}, {Leake}, {Moore}, {Nindos}, {Syntelis},
  {Yardley}, {Yurchyshyn}, \& {Zhang}}]{Patsourakos2020}
{Patsourakos}, S., {Vourlidas}, A., {T{\"o}r{\"o}k}, T., {et~al.} 2020, \ssr,
  216, 131, \dodoi{10.1007/s11214-020-00757-9}

\bibitem[{{Pesnell} {et~al.}(2012){Pesnell}, {Thompson}, \&
  {Chamberlin}}]{Pesnell2012}
{Pesnell}, W.~D., {Thompson}, B.~J., \& {Chamberlin}, P.~C. 2012, \solphys,
  275, 3, \dodoi{10.1007/s11207-011-9841-3}

\bibitem[{{Prasad} {et~al.}(2020){Prasad}, {Dissauer}, {Hu}, {Bhattacharyya},
  {Veronig}, {Kumar}, \& {Joshi}}]{Prasad2020}
{Prasad}, A., {Dissauer}, K., {Hu}, Q., {et~al.} 2020, \apj, 903, 129,
  \dodoi{10.3847/1538-4357/abb8d2}

\bibitem[{{Priest}(2014)}]{Priest2014}
{Priest}, E. 2014, {Magnetohydrodynamics of the Sun},
  \dodoi{10.1017/CBO9781139020732}

\bibitem[{{Priest} \& {D{\'e}moulin}(1995)}]{Priest1995}
{Priest}, E.~R., \& {D{\'e}moulin}, P. 1995, \jgr, 100, 23443,
  \dodoi{10.1029/95JA02740}

\bibitem[{{Priest} \& {Forbes}(2002)}]{Priest2002}
{Priest}, E.~R., \& {Forbes}, T.~G. 2002, \aapr, 10, 313,
  \dodoi{10.1007/s001590100013}

\bibitem[{{Qiu} {et~al.}(2020){Qiu}, {Guo}, {Ding}, \& {Zhong}}]{Qiu2020}
{Qiu}, Y., {Guo}, Y., {Ding}, M., \& {Zhong}, Z. 2020, \apj, 901, 13,
  \dodoi{10.3847/1538-4357/abae5b}

\bibitem[{{Rust} \& {Kumar}(1996)}]{Rust1996}
{Rust}, D.~M., \& {Kumar}, A. 1996, \apjl, 464, L199, \dodoi{10.1086/310118}

\bibitem[{{Sahu} {et~al.}(2020){Sahu}, {Joshi}, {Mitra}, {Veronig}, \&
  {Yurchyshyn}}]{Sahu2020}
{Sahu}, S., {Joshi}, B., {Mitra}, P.~K., {Veronig}, A.~M., \& {Yurchyshyn}, V.
  2020, \apj, 897, 157, \dodoi{10.3847/1538-4357/ab962b}

\bibitem[{{Sarkar} \& {Srivastava}(2018)}]{Sarkar2018}
{Sarkar}, R., \& {Srivastava}, N. 2018, \solphys, 293, 16,
  \dodoi{10.1007/s11207-017-1235-8}

\bibitem[{{Schou} {et~al.}(2012){Schou}, {Scherrer}, {Bush}, {Wachter},
  {Couvidat}, {Rabello-Soares}, {Bogart}, {Hoeksema}, {Liu}, {Duvall}, {Akin},
  {Allard}, {Miles}, {Rairden}, {Shine}, {Tarbell}, {Title}, {Wolfson},
  {Elmore}, {Norton}, \& {Tomczyk}}]{Schou2012}
{Schou}, J., {Scherrer}, P.~H., {Bush}, R.~I., {et~al.} 2012, \solphys, 275,
  229, \dodoi{10.1007/s11207-011-9842-2}

\bibitem[{{Sturrock}(1966)}]{Sturrock1966}
{Sturrock}, P.~A. 1966, \nat, 211, 695, \dodoi{10.1038/211695a0}

\bibitem[{{Sui} {et~al.}(2006){Sui}, {Holman}, \& {Dennis}}]{Sui2006}
{Sui}, L., {Holman}, G.~D., \& {Dennis}, B.~R. 2006, \apj, 646, 605,
  \dodoi{10.1086/504885}

\bibitem[{{\v{S}}vestka \& Cliver(1992)}]{Svestka1992}
{\v{S}}vestka, Z., \& Cliver, E.~W. 1992, in Eruptive Solar Flares, ed.
  Z.~{\v{S}}vestka, B.~V. Jackson, \& M.~E. Machado (Berlin, Heidelberg:
  Springer Berlin Heidelberg), 1--11

\bibitem[{{Takano} {et~al.}(1997){Takano}, {Nakajima}, {Enome}, {Shibasaki},
  {Nishio}, {Hanaoka}, {Shiomi}, {Sekiguchi}, {Kawashima}, {Bushimata},
  {Shinohara}, {Torii}, {Fujiki}, \& {Irimajiri}}]{Takano1997}
{Takano}, T., {Nakajima}, H., {Enome}, S., {et~al.} 1997, {An Upgrade of
  Nobeyama Radioheliograph to a Dual-Frequency (17 and 34 GHz) System}, ed.
  G.~{Trottet}, Vol. 483, 183, \dodoi{10.1007/BFb0106457}

\bibitem[{{Tian} {et~al.}(2018){Tian}, {Shen}, \& {Liu}}]{Tian2018}
{Tian}, Z., {Shen}, Y., \& {Liu}, Y. 2018, \na, 65, 7,
  \dodoi{10.1016/j.newast.2018.05.005}

\bibitem[{{Titov} {et~al.}(2002){Titov}, {Hornig}, \&
  {D{\'e}moulin}}]{Titov2002}
{Titov}, V.~S., {Hornig}, G., \& {D{\'e}moulin}, P. 2002, Journal of
  Geophysical Research (Space Physics), 107, 1164, \dodoi{10.1029/2001JA000278}

\bibitem[{{T{\"o}r{\"o}k} {et~al.}(2004){T{\"o}r{\"o}k}, {Kliem}, \&
  {Titov}}]{Torok2004}
{T{\"o}r{\"o}k}, T., {Kliem}, B., \& {Titov}, V.~S. 2004, \aap, 413, L27,
  \dodoi{10.1051/0004-6361:20031691}

\bibitem[{{Tsuneta} {et~al.}(1992){Tsuneta}, {Hara}, {Shimizu}, {Acton},
  {Strong}, {Hudson}, \& {Ogawara}}]{Tsuneta1992}
{Tsuneta}, S., {Hara}, H., {Shimizu}, T., {et~al.} 1992, \pasj, 44, L63

\bibitem[{{Veronig} {et~al.}(2006){Veronig}, {Karlick{\'y}}, {Vr{\v{s}}nak},
  {Temmer}, {Magdaleni{\'c}}, {Dennis}, {Otruba}, \& {P{\"o}tzi}}]{Veronig2006}
{Veronig}, A.~M., {Karlick{\'y}}, M., {Vr{\v{s}}nak}, B., {et~al.} 2006, \aap,
  446, 675, \dodoi{10.1051/0004-6361:20053112}

\bibitem[{{Veronig} {et~al.}(2018){Veronig}, {Podladchikova}, {Dissauer},
  {Temmer}, {Seaton}, {Long}, {Guo}, {Vr{\v{s}}nak}, {Harra}, \&
  {Kliem}}]{Veronig2018}
{Veronig}, A.~M., {Podladchikova}, T., {Dissauer}, K., {et~al.} 2018, \apj,
  868, 107, \dodoi{10.3847/1538-4357/aaeac5}

\bibitem[{{Vr{\v{s}}nak}(2016)}]{Vrsnak2016}
{Vr{\v{s}}nak}, B. 2016, Astronomische Nachrichten, 337, 1002,
  \dodoi{10.1002/asna.201612424}

\bibitem[{{Wang} {et~al.}(2017){Wang}, {Liu}, {Wang}, {Liu}, {Chen}, {Liu},
  {Zhou}, \& {Zhang}}]{Wang2017}
{Wang}, D., {Liu}, R., {Wang}, Y., {et~al.} 2017, \apjl, 843, L9,
  \dodoi{10.3847/2041-8213/aa79f0}

\bibitem[{{Wang} {et~al.}(2018){Wang}, {Su}, {Shen}, {Yang}, {Cao}, \&
  {Ji}}]{Wang2018}
{Wang}, Y., {Su}, Y., {Shen}, J., {et~al.} 2018, \apj, 859, 148,
  \dodoi{10.3847/1538-4357/aac0f7}

\bibitem[{{Wiegelmann} {et~al.}(2012){Wiegelmann}, {Thalmann}, {Inhester},
  {Tadesse}, {Sun}, \& {Hoeksema}}]{Wiegelmann2012}
{Wiegelmann}, T., {Thalmann}, J.~K., {Inhester}, B., {et~al.} 2012, \solphys,
  281, 37, \dodoi{10.1007/s11207-012-9966-z}

\bibitem[{{Wiegelmann, T.} \& {Inhester, B.}(2010)}]{Wiegelmann2010}
{Wiegelmann, T.}, \& {Inhester, B.} 2010, A\&A, 516, A107,
  \dodoi{10.1051/0004-6361/201014391}

\bibitem[{{Yang} {et~al.}(2015){Yang}, {Guo}, \& {Ding}}]{Yang2015}
{Yang}, K., {Guo}, Y., \& {Ding}, M.~D. 2015, \apj, 806, 171,
  \dodoi{10.1088/0004-637X/806/2/171}

\bibitem[{{Yashiro} {et~al.}(2005){Yashiro}, {Gopalswamy}, {Akiyama},
  {Michalek}, \& {Howard}}]{Yashiro2005}
{Yashiro}, S., {Gopalswamy}, N., {Akiyama}, S., {Michalek}, G., \& {Howard},
  R.~A. 2005, Journal of Geophysical Research (Space Physics), 110, A12S05,
  \dodoi{10.1029/2005JA011151}

\bibitem[{{Zhang} {et~al.}(2012){Zhang}, {Cheng}, \& {Ding}}]{Zhang2012}
{Zhang}, J., {Cheng}, X., \& {Ding}, M.-D. 2012, Nature Communications, 3, 747,
  \dodoi{10.1038/ncomms1753}

\bibitem[{{Zheng} {et~al.}(2019){Zheng}, {Yang}, {Rao}, {Liu}, {Zhong}, {Wang},
  {Song}, {Li}, \& {Chen}}]{Zheng2019}
{Zheng}, R., {Yang}, S., {Rao}, C., {et~al.} 2019, \apj, 875, 71,
  \dodoi{10.3847/1538-4357/ab0f3f}

\bibitem[{{Zhong} {et~al.}(2021){Zhong}, {Guo}, \& {Ding}}]{Zhong2021}
{Zhong}, Z., {Guo}, Y., \& {Ding}, M.~D. 2021, Nature Communications, 12, 2734,
  \dodoi{10.1038/s41467-021-23037-8}

\bibitem[{{Zuccarello} {et~al.}(2015){Zuccarello}, {Aulanier}, \&
  {Gilchrist}}]{Zuccarello2015}
{Zuccarello}, F.~P., {Aulanier}, G., \& {Gilchrist}, S.~A. 2015, \apj, 814,
  126, \dodoi{10.1088/0004-637X/814/2/126}

\end{thebibliography}
\end{document}